\documentclass[aps,prd,preprint,showpacs,showkeys,nofootinbib,superscriptaddress]{revtex4-1}
\pdfoutput=1
\usepackage{amsmath}
\usepackage{amssymb}
\usepackage{dcolumn}% Align table columns on decimal point
\usepackage{bm}% bold math
\usepackage{multirow}
\usepackage{blkarray}
\usepackage{feynmp}                                     
\usepackage{hyperref,graphicx}           % usepackage without driver option
\usepackage{caption}
\usepackage{subcaption}
\usepackage{xcolor}
\usepackage[utf8]{inputenc}

\begin{document}

\title{Singlet Scalar Top Partners\\ from Accidental Supersymmetry}

\author{Hsin-Chia Cheng}
\affiliation{Center for Quantum Mathematics and Physics (QMAP), Department of Physics,\\ University of California, Davis, California, 95616, USA}
\affiliation{School of Natural Sciences, Institute for Advanced Study, Princeton, New Jersey 08540, USA}
\author{Lingfeng Li}
\affiliation{Center for Quantum Mathematics and Physics (QMAP), Department of Physics,\\ University of California, Davis, California, 95616, USA}
\author{Ennio Salvioni}
\affiliation{Physik-Department, Technische Universit\"at M\"unchen, 85748 Garching, Germany}
\author{Christopher B. Verhaaren}\email[Email: ]{cheng@physics.ucdavis.edu}\email{llfli@ucdavis.edu}\email{ennio.salvioni@tum.de}\email{cbverhaaren@ucdavis.edu}
\affiliation{Center for Quantum Mathematics and Physics (QMAP), Department of Physics,\\ University of California, Davis, California, 95616, USA}

%\date{\today}

\begin{abstract}
We present a model wherein the Higgs mass is protected from the quadratic one-loop top quark corrections by scalar particles that are complete singlets under the Standard Model (SM) gauge group. While bearing some similarity to Folded Supersymmetry, the construction is purely four dimensional and enjoys more parametric freedom, allowing electroweak symmetry breaking to occur easily. The cancelation of the top loop quadratic divergence is ensured by a $Z_3$ symmetry that relates the SM top sector and two hidden top sectors, each charged under its own hidden color group. In addition to the singlet scalars, the hidden sectors contain electroweak-charged supermultiplets below the TeV scale, which provide the main access to this model at colliders. The phenomenology presents both differences and similarities with respect to other realizations of neutral naturalness. Generally, the glueballs of hidden color have longer decay lengths. The production of hidden sector particles results in quirk or squirk bound states, which later annihilate. We survey the possible signatures and corresponding experimental constraints.
\end{abstract}
%\preprint{arXiv:1803.03651~[hep-ph]}
\preprint{TUM-HEP-1134-18}

\maketitle
\tableofcontents
%\clearpage

\section{Introduction\label{sec.intro}}
The discovery of the Higgs boson~\cite{Aad:2012tfa,Chatrchyan:2012ufa} at the Large Hadron Collider (LHC) successfully completes the Standard Model (SM) but also accentuates its mysteries. In particular, the well-known hierarchy between the mass of the Higgs boson and the Planck scale becomes ever more puzzling as the LHC acquires new data. No colored top partner, that could cancel the quadratic top loop contribution to the Higgs mass, has been found below the TeV scale. Besides some special cases which may evade the strong experimental bounds, this implies a serious fine tuning of the Higgs mass parameter.

These bounds have largely driven the recent interest in neutral naturalness (NN), that is, symmetry-based solutions to the hierarchy problem where the symmetry partners of quarks are not charged under SM color. The twin Higgs framework~\cite{Chacko:2005pe} led the way with fermionic top partners that are complete SM singlets. This involves a new, ``twin,'' sector related to the SM by a discrete $Z_2$ symmetry, but charged under distinct gauge groups. The $Z_2$ guarantees the equality of parameters necessary to cancel the leading one-loop quantum corrections to the Higgs mass. 

Scalar NN top partners first appeared in Folded Supersymmetry (FSUSY)~\cite{Burdman:2006tz}. The realistic model is formulated in an orbifold extra dimension, with the gauge group extended to $SU(3)^2 \times SU(2) \times U(1)$. The matter content is doubled from that of the minimal supersymmetric SM (MSSM), with each copy having its own $SU(3)$ gauge group, but sharing a single electroweak (EW) structure.  A $Z_2$ symmetry relates the SM quark and lepton multiplets to the folded sector, as well as the two $SU(3)$ gauge groups. SUSY is broken by the Scherk-Schwarz~\cite{Scherk:1978ta,Scherk:1979zr} mechanism such that within the matter multiplets only the SM fermions and folded sector scalars have zero modes. The true top squarks (stops), which carry SM color, have masses set by the compactification scale, assumed to be multi-TeV. However, due to an accidental SUSY, the quadratic top loop contribution to the Higgs potential is cut off by the folded stops, which only receive masses at one loop and remain light. The folded stops are SM color neutral, so they are not subject to the strong experimental bounds on colored particles. They do carry SM EW quantum numbers, which govern their collider phenomenology. 

In the original FSUSY the cancelation of the top loop contribution to the Higgs mass term is in fact too effective, which makes EW symmetry breaking difficult. This can be fixed by some modifications of the model. For instance, Ref.~\cite{Cohen:2015gaa} considered twisting the boundary conditions of the Scherk-Schwarz mechanism by an $SU(2)_R$ rotation of phase $\alpha$. For $\alpha=0$, $\mathcal{N}=1$ SUSY is preserved and there is no zero mode in the folded sector, while the original FSUSY corresponds to $\alpha=1/2$. For $0 < \alpha < 1/2$ the folded stops have tree-level masses and the cancelation of the one-loop top contribution to the Higgs mass is incomplete, which provides parameter regions with the correct EW symmetry breaking. 

One may ask how essential the extra dimensional setup is for color-neutral scalar top partners, and whether it is possible to construct a four-dimensional (4D) mimic of FSUSY. A na\"ive attempt is to write down a 4D SUSY model with SUSY-breaking terms which reproduce the low-lying spectrum of FSUSY. For example, consider the following superpotential for the top sector
\begin{equation}
W_{Z_2} = y_t \left(Q H u^c + Q_f H u_f^c\right) + M\left( Q_f Q'^c_f + u'_f u^c_f \right) ,
\end{equation}
where $Q = (t, b)^T, u^c$ are SM quarks and $H$ is the (up-type) Higgs, whereas the ``$f$'' subscript denotes hidden sector fields, plus the soft SUSY-breaking masses
\begin{equation}
V_{\rm s} = \widetilde{m}^2\left(| \widetilde{Q}|^2 + \left| \tilde{u}^c\right|^2\right) - \widetilde{m}^2\left(|  \widetilde{Q}_f|^2+\left| \tilde{u}_f^c\right|^2\right) .
\end{equation}
When $\widetilde{m} \to M$, the true stops $\tilde{t}, \tilde{u}^c$ become heavy while the folded stops $\tilde{t}_f, \tilde{u}^c_f$ become light, which simulates the truncated FSUSY spectrum. However, the coupling of the light color-neutral scalars to the Higgs is given by $y_t^2(1-M^2/\widetilde{m}^2)$ and vanishes in the limit $\widetilde{m} = M$. As a result, there is no accidental SUSY and the light scalars cannot cancel the top loop contribution to the Higgs potential, leaving the Higgs mass-squared quadratically sensitive to the heavy mass scale $M$.

This problem can be solved by further extending the hidden sector. In this paper we construct a 4D model based on the gauge group $SU(3)^3 \times SU(2) \times U(1)$, where the contribution to the Higgs potential from the top sector is calculable and finite, and only has logarithmic sensitivity to the heavy mass scale $M$ of the true stops. The 4D construction allows more freedom compared to 5D models. The superpartners of the SM EW sector are not governed by the same mass scale as the true stops, and can still be light. We do not need hidden sector partners for the first two generations, since their contributions to the Higgs potential are small. Furthermore, freedom in hidden sector hypercharge assignments allows the top partners to be fully SM-singlet scalars, even though some new EW-charged particles are needed for the complete model.

Detecting completely SM singlet scalar top partners provides an interesting experimental challenge. While the Higgs may couple to new states beyond the SM (BSM), its couplings to SM fields can be very SM-like. The collider phenomenology of such scenarios has been explored~\cite{Craig:2013xia,Curtin:2015bka} using bottom-up simplified approaches, but a complete model has not yet appeared.\footnote{An independent approach has been pursued in Ref.~\cite{Cohen:2018mgv}.} The framework outlined in this work, however, has phenomenology determined largely by the EW-charged particles that accompany the SM singlet top partners. 

In the next section we describe the model and discuss the structure of the Higgs potential it generates. Section~\ref{s.construct} explores a possible mechanism for obtaining the special soft mass structure required by our construction, with additional details provided in Appendix~\ref{s.softmasses}. The phenomenology of the model and the constraints on its parameters are analyzed in Sec.~\ref{sec:pheno}, while technical derivations of important results are given in Appendices~\ref{a.lambdaQCD}, \ref{a.QuirkoniumMixing}, and \ref{app:squirkonium}. We conclude in Sec.~\ref{sec:conclusion}.

\section{A Tripled Top Model\label{sec.model}}
We extend a supersymmetric SM by adding two copies of a ``hidden'' top quark sector, which we label $B$ and $C$, with $A$ labeling the SM sector. The hidden tops are not charged under the SM color but carry hidden colors of $SU(3)_B$ and $SU(3)_C$ respectively. Both $SU(2)$ doublet and singlet hidden tops have mirror partners and form vector-like pairs. The superpotential of the three top sectors takes the form
\begin{align}
W_{Z_3} & = y_t\left( Q_A H u^c_A + Q_B H u^c_B + Q_C H u^c_C \right)+ M  ( u^\prime_B  u_B^c + u^\prime_C u_C^c ) +  \omega (Q_B Q_B^{\prime c} + Q_C Q_C^{\prime c} )  \,.
\end{align}
The couplings with the (up-type) Higgs respect a $Z_3$ symmetry, which also relates the three $SU(3)$ gauge groups. Accordingly, we call this a ``tripled top'' framework. The supersymmetric vector-like mass terms $M$ and $\omega$ of the hidden sectors softly break $Z_3$ to $Z_2$. $M$ is taken to be multi-TeV while $\omega$ is assumed to be below 1 TeV, which we will see keeps the Higgs mass light.

The SM fields have the usual charges under the EW $SU(2)_L \times U(1)_Y$,
\begin{equation}
H = \begin{pmatrix} h^+ \\ h^0 \end{pmatrix} \sim \mathbf{2}_{1/2}\,, \qquad Q_{A} = \begin{pmatrix} t_{A} \\ b_{A} \end{pmatrix} \sim \mathbf{2}_{1/6}\,, \qquad u_A^c \sim \mathbf{1}_{-2/3}\,,
\end{equation}
which also defines the component fields. For the $B$ and $C$ sectors we choose
\begin{equation}
Q_{B,C} = \begin{pmatrix} t_{B,C} \\ b_{B,C} \end{pmatrix} \sim \mathbf{2}_{- 1/2}\,, \qquad Q^{\prime c}_{B,C} = \begin{pmatrix} b^{ \prime c}_{B,C} \\ t^{ \prime c}_{B,C} \end{pmatrix} \sim \mathbf{2}_{1/2}\,,\qquad u^c_{B,C},\, u^\prime_{B,C} \sim \mathbf{1}_0\,.
\end{equation}
Notice that ``$u$'' fields are $SU(2)_L$ singlets, while ``$t$'' states belong to doublets. We emphasize that the hypercharge assignments of the $B$ and $C$ fields are free, up to keeping the Yukawa terms gauge invariant. We specifically choose the $SU(2)_L$ singlets to be complete SM singlets. Their scalar components play the roles of top partners, cutting off the quadratic top loop contribution to the Higgs potential.

The leading soft SUSY-breaking masses are assumed to take the form
\begin{equation}
V_{\rm s} = \widetilde{m}^2 \left( | \widetilde{Q}_A |^2 + \left| \tilde{u}^{c}_A  \right|^2 \right) - \widetilde{m}^2 \left( \left| \tilde{u}_B^c \right|^2 + \left| \tilde{u}_{C}^c \right|^2 \right) \,.\label{e.softmass}
\end{equation}
The opposite-sign, equal-magnitude soft mass terms ensure the Higgs potential from the top sectors is calculable and finite. Their possible origins are discussed in the next section. The soft SUSY-breaking masses raise the colored stop masses and lower the masses of $\tilde{u}_B^c$ and $\tilde{u}_C^c$. To make $\tilde{u}_B^c$ and $\tilde{u}_C^c$ light $\widetilde{m}$ must be close to $M$, so the masses of $\tilde{u}_B^c$, $\tilde{u}_C^c$ are, before mixing effects from the Higgs vacuum expectation value (VEV), given by 
\begin{equation}
\Delta \equiv \sqrt{M^2 - \widetilde{m}^2}\, \ll  M.
\end{equation}
At the same time the colored stop masses are raised to the multi-TeV scale. 

In addition to the soft SUSY breaking masses in Eq.~(\ref{e.softmass}), the $A$ sector gluino and light generation squarks must also have multi-TeV SUSY-breaking masses to evade LHC bounds. All other fields can receive subleading SUSY-breaking masses of a few hundred GeV which split the fermions and bosons in the supermultiplets, without spoiling naturalness. %Since they do not generate large corrections to the Higgs potential, they do not cause any naturalness problem.

\subsection{Higgs potential}
We now demonstrate that the one-loop quadratic contribution to the Higgs potential from the top quark is canceled by the neutral top partners. Consequently, the Higgs potential has no quadratic dependence on the heavy scale $M$. Before deriving the complete expression of one-loop Coleman-Weinberg (CW)~\cite{Coleman:1973jx} potential for general parameters, we show the protection of the Higgs mass in the limit $\widetilde{m} \to M\,(\Delta \to 0)$. This case is similar to the original FSUSY, whose authors pointed out that the cancelation of divergences is tied to the apparent supersymmetric structure of the theory when \emph{only scalar} labels are exchanged. The Higgs mass is protected by this accidental supersymmetry. 

\begin{figure}[t]
\begin{center}
\includegraphics[width=9.5cm]{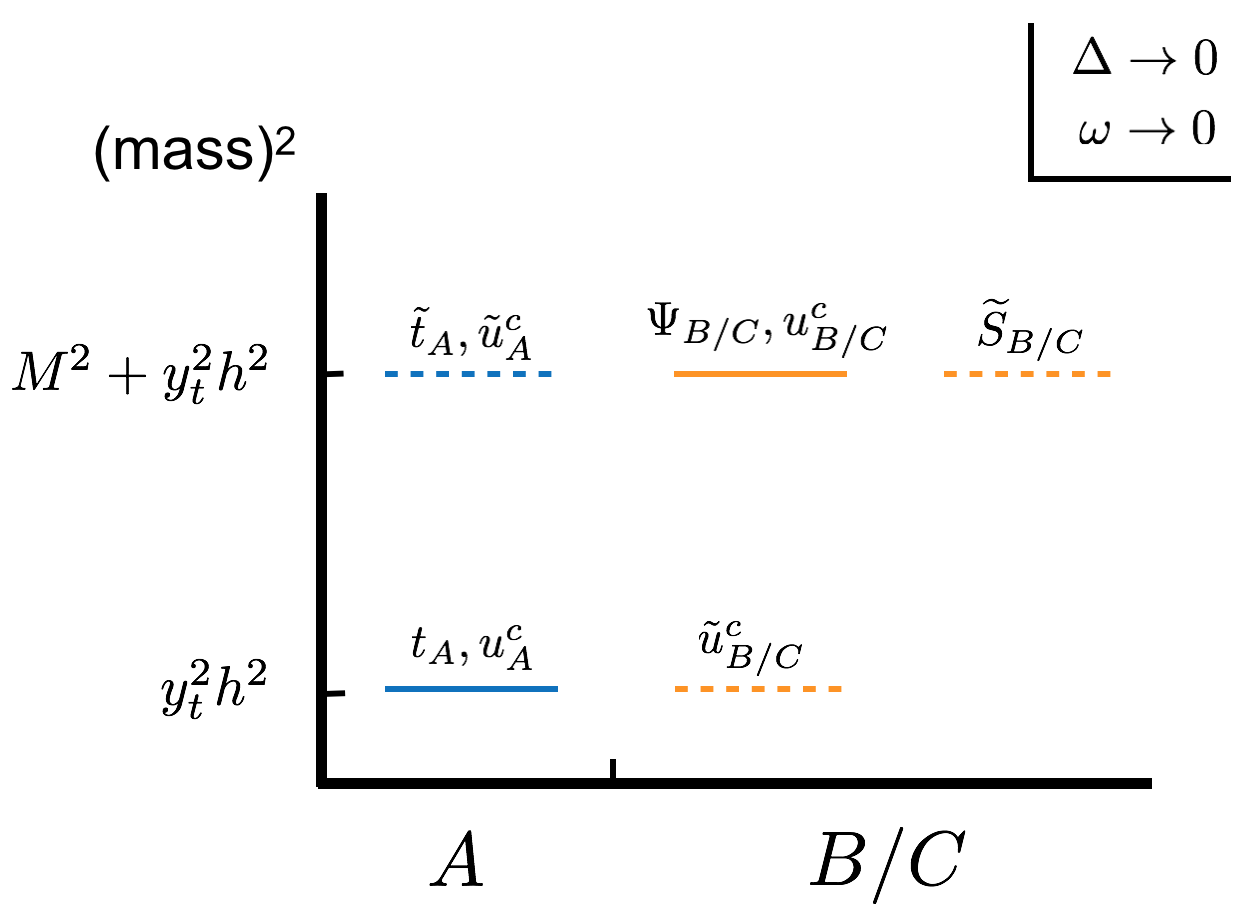}
\end{center}
\caption{Mass spectrum in the limit $\Delta \to 0$ ($\widetilde{m} \to M$) and $\omega \to 0$, illustrating the accidental SUSY that protects the Higgs mass. We have $\Psi_{B/C} = \cos\theta_L u^\prime_{B/C} - \sin \theta_L t_{B/C}$ and $\widetilde{S}_{B/C} = \cos\theta_L \tilde{u}_{B/C}^\prime - \sin \theta_L \tilde{t}_{B/C}$, where $ \left. \sin\theta_L \right|_{\omega\, \to \, 0} = - y_t h / \sqrt{M^2 + y_t^2 h^2}\,$.}
\label{f:cancellationsketch}
\end{figure}

The Higgs-dependent scalar masses that arise from the superpotential of the top sector can be divided into five groups
\begin{align}
V_1 \,=&\; y_t^2 h^2 \left( |\tilde{t}_A|^2 + | \tilde{u}_A^c |^2 +  |\tilde{t}_B|^2 + | \tilde{u}_B^c |^2 +  |\tilde{t}_C|^2 + | \tilde{u}_C^c |^2 \right), \nonumber \\
V_2 \,=&\; M^2 \left(| \tilde{u}_B^c |^2 + | \tilde{u}_C^c |^2\right) + M^2 \left( | \tilde{u}^\prime_B |^2 + | \tilde{u}^\prime_C |^2 \right), \nonumber \\
V_3 \,=&\; y_t h M \left(\tilde{t}_B^\ast \tilde{u}^\prime_B + \tilde{t}_C^\ast \tilde{u}^\prime_C + \mathrm{h.c.}\right), \\
V_4 \,=&\; \omega^2 \left(  |\tilde{t}_B|^2 + | \tilde{t}_B^{\, \prime c} |^2 + |\tilde{t}_C|^2 + | \tilde{t}_C^{\, \prime c} |^2 \right), \nonumber \\
V_5 \,=&\; y_t h \,\omega \left( \tilde{u}_B^{c\,\ast} \tilde{t}_B^{\,\prime c} +  \tilde{u}_C^{c\,\ast} \tilde{t}_C^{\,\prime c} + \mathrm{h.c.}\right), \nonumber
\end{align}
where we defined $h \equiv \mathrm{Re} \,h^0$. After the soft masses 
\begin{equation}
V_{\rm soft} (\widetilde{m}) =  \widetilde{m}^2 \left( | \tilde{t}_A |^2 + \left| \tilde{u}^{c}_A  \right|^2 \right) - \widetilde{m}^2 \left( | \tilde{u}_B^c |^2 + | \tilde{u}_{C}^c |^2 \right)
\end{equation}
with $\widetilde{m}=M$ are added, we expect that $\tilde{u}_B^c$, $\tilde{u}_C^c$ will replace the roles of $\tilde{t}_A$ and $\tilde{u}_A^c$ as they become massless.
Thus, the exchange
\begin{equation}
\tilde{t}_A \leftrightarrow\, \tilde{u}_B^c, \ \ \ \ \,\tilde{u}_A^c\, \leftrightarrow\, \tilde{u}_C^c,
\end{equation}
denoted by the mapping $\tau$, may lead to an accidental SUSY. Notice first that under this exchange $V_1$, $V_3$, and $V_4$ are invariant, that is $\tau[V_{1,3,4}]=V_{1,3,4}$. Next, in the limit $\widetilde{m}=M$
\begin{equation}
\tau\left[V_2+V_{\rm soft} (M) \right]=V_2\,,
\end{equation}
which is a completely supersymmetric scalar potential. In other words, except for $V_5$, in the limit $\widetilde{m}\to M$ there is an accidental supersymmetry, a potential with no SUSY-breaking terms. In this limit we expect only $V_5$ to contribute to the Higgs mass, yielding $\sim N_c y_t^2 \omega^2 \ln M^2 / (16 \pi^2)$ at one loop. As long as $\omega$ is only a few hundred GeV, it does not cause a naturalness problem. Figure \ref{f:cancellationsketch} shows the spectrum in the limit $\widetilde{m}\to M$ and $\omega\to 0\,$: each fermion is exactly degenerate with two scalars of equal coupling to the Higgs, ensuring that the Higgs mass parameter vanishes exactly.

This expectation is borne out by the explicit CW computation (for general $\widetilde{m} \neq M$). First, we find that there are neither quadratic nor logarithmic divergences, since both $\mathrm{STr}\, \mathcal{M}^2 = 0$ and $\mathrm{STr}\, \mathcal{M}^4 = - 8 N_c \,\Delta^2 (M^2 - \Delta^2)$ are field-independent. Proceeding to the finite pieces, we find the mass term
\begin{align} \label{eq:hmassgeneral}
V_{h^2} = &- \,\frac{N_c y_t^2 h^2}{8\pi^2}\left[-\left(M^2-\Delta^2 \right)\ln\left(1-\frac{\Delta^2}{M^2} \right)-\frac{\Delta^4}{\omega^2-\Delta^2}\ln\frac{M^2}{\Delta^2}\right.\nonumber\\
&\left.\quad\quad\quad\quad\quad+\frac{\omega^4(M^2-\Delta^2)}{(M^2-\omega^2)(\omega^2-\Delta^2)}\ln\frac{M^2}{\omega^2} \right] \\[12pt]
\approx &- \,\frac{N_c y_t^2 h^2}{8\pi^2}\left[\frac{\omega^4}{\omega^2-\Delta^2}\ln\frac{M^2}{\omega^2}-\frac{\Delta^4}{\omega^2-\Delta^2}\ln\frac{M^2}{\Delta^2} + \Delta^2 \right]+ O \left(M^{-2} \right). \label{eq:hmassgeneralapprox}
\end{align}
Notice that $V_{h^2}$ scales as $\sim \omega^2 \ln M^2 /(16\pi^2)$ when $\Delta \to 0$, as expected. Hence, the Higgs mass is set by $\omega$, the scale of the electroweak doublets, and $\Delta$, the scale of the singlets, and is only logarithmically sensitive to $M$, the scale of colored states. We also calculate the quartic coupling,
\begin{align} \label{eq:quartictotalnewnot}
V_{h^4}=&\;\frac{N_c y_t^4 h^4}{16 \pi^2}\left\{ \frac32+\frac{2\omega^2(M^2-\Delta^2)(M^2\Delta^2-\omega^4)}{(M^2-\omega^2)^2(\omega^2-\Delta^2)^2}+\ln\frac{M^2}{y_t^2h^2}+\ln\left(1-\frac{\Delta^2}{M^2} \right) \right.\nonumber\\
&\left. +\frac{\Delta^4(\Delta^2-3\omega^2)}{(\omega^2-\Delta^2)^3}\ln\frac{M^2}{\Delta^2}
-\left[\frac{\omega^4(\omega^2-3M^2)}{(M^2-\omega^2)^3}+\frac{\omega^4(\omega^2-3\Delta^2)}{(\omega^2-\Delta^2)^3} \right]\ln\frac{M^2}{\omega^2}\right\} \\[12pt]
\approx&\, \; \frac{N_c y_t^4 h^4}{16 \pi^2} \left\{ \frac32 +\frac{2\omega^2\Delta^2}{(\omega^2-\Delta^2)^2}+ \ln \frac{M^2}{y_t^2 h^2} +\frac{\Delta^4(\Delta^2-3\omega^2)}{(\omega^2-\Delta^2)^3}\ln\frac{M^2}{\Delta^2}  \right.\nonumber\\
&\left. -\frac{\omega^4(\omega^2-3\Delta^2)}{(\omega^2-\Delta^2)^3}\ln\frac{M^2}{\omega^2} \right\} + O(M^{-2}). \label{eq:quartictotalnewnotapprox}
\end{align}
We see that for $\Delta \to 0$ and at leading order in $\omega^2/M^2 \ll 1$,
\begin{equation}
V_{h^4} \simeq \frac{N_c y_t^4 h^4}{16 \pi^2} \left(\frac32 + \ln \frac{\omega^2}{y_t^2 h^2}\right),
\end{equation}
which is independent of $M$.

To get some feeling for the numbers, recall that in the SM the parameters of the Higgs potential,
\begin{equation}
V = m^2 h^2 + \lambda h^4,
\end{equation}
take the values $m^2_{\rm SM} \simeq -(88\;\mathrm{GeV})^2$ and $\lambda_{\rm SM} \simeq 0.13$, which yield the VEV $v= \sqrt{2}\, \langle h \rangle = \sqrt{-m_\text{SM}^2/\lambda_{\rm SM}} \simeq 246$ GeV and physical Higgs boson mass $m_h= \sqrt{2 \lambda_\text{SM}}\, v \simeq 125\; \text{GeV}$. From Eqs.~\eqref{eq:hmassgeneral} and \eqref{eq:quartictotalnewnot}, setting for example $M = 2$ TeV, $\omega = 500$ GeV, $\Delta = 300$ GeV and neglecting the running of $y_t$, we find $ m^2 \simeq - (196\;\mathrm{GeV})^2$ and $\lambda \simeq 0.073\,$. The gauge contribution to the Higgs potential gives an additional quartic $\leq (g^2 + g^{\prime \,2})/8 = m_Z^2/ (2v^2) \simeq 0.069$, where the maximal value is attained for $\tan\beta \to \infty$. Thus, the total Higgs quartic is in the correct range to obtain $m_h \simeq 125\; \text{GeV}$, although a precise assessment requires the inclusion of the leading two-loop corrections, which goes beyond the scope of this paper. If necessary, additional contributions to the Higgs quartic may arise as in the next-to-minimal supersymmetric SM, by adding a singlet $S$ to the model with a superpotential term $\lambda S H_u H_d$. Notice also that the leading order expression of the quartic in Eq.~\eqref{eq:quartictotalnewnotapprox} is symmetric under $\Delta \leftrightarrow \omega$, and the leading order expression of the mass in Eq.~\eqref{eq:hmassgeneralapprox} is almost symmetric. Therefore, the Higgs potential does not prefer a specific hierarchy between $\Delta$ and $\omega$.

While the Higgs potential has no quadratic sensitivity to the heavy colored stop mass scale $M$, it requires the particular form of soft masses in Eq.~\eqref{e.softmass} and $\widetilde{m}$ close to $M$, both of which may be new sources of tuning. The possible origin of the structure of the soft masses is discussed in the next section. The requirement $\Delta = \sqrt{M^2-\widetilde{m}^2} \ll M$ corresponds to a tuning of $\sim\Delta^2/M^2$ without a physical explanation that relates the supersymmetric mass $M$ and the SUSY-breaking mass $\widetilde{m}$. For $M=2$~TeV, $\Delta = 300$~GeV we have $\Delta^2/M^2 \sim 2\%$, which is no worse than most currently surviving models. This may be improved if a dynamical mechanism that relates $M\sim \widetilde{m}$ can be identified. If one uses the fine tuning measure defined by Barbieri and Giudice~\cite{Barbieri:1987fn}, then the tuning of the Higgs mass relative to $\Delta$ and $\omega$ also needs to be considered. It may seem, after multiplying these two tunings, that not much is gained compared to an MSSM with heavy stops, other than removing the large logarithm. Similar situations also occur in other models with ``double protections,'' such as SUSY twin Higgs or SUSY little Higgs~\cite{Craig:2013fga,Katz:2016wtw,Birkedal:2004xi,Chankowski:2004mq,Berezhiani:2005pb,Roy:2005hg,Csaki:2005fc,Bellazzini:2008zy,Bellazzini:2009ix,Falkowski:2006qq,Chang:2006ra}. Na\"ively, a natural attempt to alleviate a large hierarchy problem is to break the hierarchy into several smaller steps, by inserting new physics at intermediate scales. Due to the chain rule, however, in this case the Barbieri-Giudice fine tuning does not improve. This is somewhat counter-intuitive and raises the question whether this definition is suitable to compare the fine-tuning of different models or of different model parameters.

It is known that the Barbieri-Giudice formula actually measures sensitivity. But sensitivity to a parameter does not always imply tuning. For example, the proton mass is very sensitive to the QCD coupling, but is certainly not considered fine-tuned. Reference~\cite{Anderson:1994dz} proposed that a better measure of fine-tuning for a set of parameters of a given model is the ratio of the sensitivity derived from that set of parameters and the ``typical'' or ``average'' sensitivity of that model. This fixes unambiguously the normalization of the measure and ensures that a typical sensitivity is not penalized, which allows for a fairer comparison of the fine-tuning of different parameter points or of different models. For instance, working in the mSUGRA scenario the authors obtained average sensitivities of $\sim 4$ for the universal scalar mass and $\sim 10$ for the gaugino mass. Thus if one obtains the same sensitivity due to the scalar mass and the gaugino mass, the tuning due to the scalar mass should be considered more severe. Notice that when a physical quantity receives two contributions of opposite sign the typical sensitivity is always bigger than one, since one of the contributions must be larger in size than the resulting observed value.

If one adopts this point of view, then dividing a large hierarchy into several steps of small hierarchies does alleviate the fine-tuning, because at each step one should divide by the average sensitivity of that step. In our case, the sensitivity of the Higgs mass to $\Delta$ and $\omega$ is similar to the sensitivity of the Higgs mass to the stop mass in the MSSM. The sensitivity of $\Delta$ to $M$ is $M^2/\Delta^2$, as mentioned earlier, and it is reasonable to assume an average sensitivity of $2\,$-$\,3$ for this second step. Then, dividing by this additional typical sensitivity and recalling that in our model the Higgs mass is not enhanced by a large logarithm, we expect to gain an overall factor of $\sim 5$ when comparing the tuning of our model to that of an MSSM with stop mass equal to $M$.
 
\section{Opposite Sign Soft Masses\label{s.construct}}
In this section we discuss possible mechanisms of obtaining soft masses for the colored stops and $\tilde{u}_B^c$ and $\tilde{u}_C^c$ which have equal magnitude but opposite sign. This particular form indicates the soft masses may be proportional to some charges where $Q_A$ and $u_A^c$ have $+1$, $u_B^c$ and $u_C^c$ have $-1$, and all other top sector fields and the Higgs have zero charge. The simplest possibility is that these leading soft masses come from a $U(1)$ $D$-term, with the various fields charged as above. However, the Yukawa terms with the Higgs field have nonzero charges, i.e., $+2$ for $Q_A H u^c_A $, $-1$ for $Q_B H u^c_B$ and  $Q_C H u^c_C$. To write down these terms one needs to insert $U(1)$ breaking VEVs and the different charges make it difficult to justify equal Yukawa couplings.

Alternatively, these charges may come from an accidental symmetry of some strong dynamics, which is not necessarily respected by the Yukawa terms if they do not arise directly from the strong dynamics. In this case, the top fields that couple to the Higgs can be composite degrees of freedom. For example, they could be the meson fields of an s-confining theory such as an $SU(N)$ gauge theory with $F=N+1$ flavors~\cite{Seiberg:1994pq}. In such a theory, the soft SUSY-breaking masses of the composite mesons are related to the soft SUSY-breaking masses of their constituent fields by anomalous $U(1)$ symmetries~\cite{ArkaniHamed:1998wc,Luty:1999qc}. In particular, for a confining group $G$ with strong scale $\Lambda_G$, under which the constituent superfields $P_i$ transform in representations $r_i$, the soft masses of the mesons $M_{ij}=P_i\overline{P}_j/\Lambda_G$ are obtained by generalizing the results in Refs.~\cite{ArkaniHamed:1998wc,Luty:1999qc} (see Appendix~\ref{s.softmasses}),
\begin{equation}
m^2_{ij}=m^2_{P_i}+m^2_{\overline{P}_j}- \frac{2 }{b}\sum_k T_{r_k}\left(m^2_{P_k}+m^2_{\overline{P}_k}\right), \label{e.mesonMass}
\end{equation}
where $m^2_{P_i}$ and $T_{r_i}$ are the squared soft mass and Dynkin index of the $i$th constituent, respectively, and $b$ is the coefficient of the gauge coupling beta function
\begin{equation}
\frac{d\phantom{\mu}}{d\ln\mu^2}\frac{1}{g^2}=\frac{b}{16\pi^2}\,. \label{e.betadef}
\end{equation}

\subsection{Example Construction}
As a concrete example consider $G=SU(2)$ gauge theories with 6 doublets ($F=3$). The Dynkin index of the fundamental representation is $T=1/2$ and the beta function coefficient is $b= 3 N - F =3 $. Each top field coupled to the Higgs is embedded in the mesons of a separate $SU(2)$ gauge theory. For the mesons to have the correct color quantum numbers, %of the $SU(3)$ in the $A, B, C$ sectors
we take the constituent quarks $P_i$ to be color triplets or anti-triplets of the corresponding sector while the constituent antiquarks $\overline{P}_i$ only carry EW charges.\footnote{For $SU(2)$ the fundamental representation and the antifundamental representation are equivalent, but for convenience we still distinguish quarks and antiquarks. Note that there are also mesons made of 2 quarks or 2 antiquarks.} As a result, the 3 constituent quarks must have the same soft mass, while the constituent antiquarks can have different soft masses unless they belong to an EW doublet. Interestingly, Eq.~\eqref{e.mesonMass} implies that if the 3 quarks and 3 antiquarks have universal soft masses (the soft masses of quarks $\widetilde{m}_P^2$ and antiquarks $\widetilde{m}^2_{\overline{P}}$ do not need to be equal), the soft masses of the mesons made of a quark and an antiquark vanish at leading order. Therefore, it is appropriate to embed $Q_{B,C}$, which should have no soft masses at leading order, as
\begin{equation}
\begin{blockarray}{lccc}
& \widetilde{m}_P^2 & \widetilde{m}_P^2 & \widetilde{m}_P^2 \\
\begin{block}{l(ccc)}
\widetilde{m}_{\overline{P}}^2\phantom{i}\phantom{i} & \BAmulticolumn{3}{c}{\multirow{2}{*}{$Q_{B,C}$}} & & \\ 
\widetilde{m}_{\overline{P}}^2\phantom{i}   & & & \\ \BAhhline{&---}
\widetilde{m}_{\overline{P}}^2\phantom{i}  & & & \\
\end{block}
\end{blockarray}\;\; .
\end{equation}
On the other hand, $u_B^c, u_C^c$ and $Q_A, u_A^c$ should have opposite soft masses and can be embedded as
\begin{equation}
\begin{blockarray}{lccc}
& \widetilde{m}_P^2 & \widetilde{m}_P^2 & \widetilde{m}_P^2 \\
\begin{block}{l(ccc)}
\widetilde{m}_{\overline{P}_1}^2\phantom{i}\phantom{i} & \BAmulticolumn{3}{c}{u^c_{B,C}} & &\\ \BAhhline{&---}
\widetilde{m}_{\overline{P}_1}^2\phantom{i}   & & & \\ 
\widetilde{m}_{\overline{P}_2}^2\phantom{i}  & & & \\
\end{block}
\end{blockarray}\;\; ,
\end{equation}
\begin{equation}
\label{e.mesonmatrix}
\begin{blockarray}{lccc}
& \widetilde{m}_P^2 & \widetilde{m}_P^2 & \widetilde{m}_P^2 \\
\begin{block}{l(ccc)}
\widetilde{m}_{\overline{P}_2}^2\phantom{i}\phantom{i} & \BAmulticolumn{3}{c}{\multirow{2}{*}{$Q_A$}} & & \\ 
\widetilde{m}_{\overline{P}_2}^2\phantom{i}   & & & \\ \BAhhline{&---}
\widetilde{m}_{\overline{P}_1}^2\phantom{i}  & & & \\
\end{block}
\end{blockarray}\;\;,\ \ \ \ \ \
\begin{blockarray}{lccc}
& \widetilde{m}_P^2 & \widetilde{m}_P^2 & \widetilde{m}_P^2 \\
\begin{block}{l(ccc)}
\widetilde{m}_{\overline{P}_2}^2\phantom{i}\phantom{i} & \BAmulticolumn{3}{c}{u^c_A} & &\\ \BAhhline{&---}
\widetilde{m}_{\overline{P}_2}^2\phantom{i}   & & & \\ 
\widetilde{m}_{\overline{P}_1}^2\phantom{i}  & & & \\
\end{block}
\end{blockarray}\;\; ,
\end{equation}
where $\widetilde{m}^2_{\overline{P}_1}\neq\widetilde{m}^2_{\overline{P}_2}$.

From the meson soft mass formula Eq.~\eqref{e.mesonMass} we find
\begin{align}
\widetilde{m}_{{Q}_{B,C}}^2=&\;\widetilde{m}_P^2+\widetilde{m}_{\overline{P}}^2-\widetilde{m}_P^2-\widetilde{m}_{\overline{P}}^2=0,\\
\widetilde{m}_{{u}^c_{B,C}}^2=&\;\widetilde{m}_P^2+\widetilde{m}_{\overline{P}_1}^2-\widetilde{m}_P^2-\frac23 \widetilde{m}_{\overline{P}_1}^2-\frac13 \widetilde{m}_{\overline{P}_2}^2=\frac{\widetilde{m}_{\overline{P}_1}^2-\widetilde{m}_{\overline{P}_2}^2}{3}, \\
\widetilde{m}^2_{Q_A,u^c_A}=&\; \widetilde{m}_P^2+\widetilde{m}_{\overline{P}_2}^2-\widetilde{m}_P^2-\frac23 \widetilde{m}_{\overline{P}_2}^2-\frac13 \widetilde{m}_{\overline{P}_1}^2= \frac{\widetilde{m}_{\overline{P}_2}^2-\widetilde{m}_{\overline{P}_1}^2}{3}=-\,\widetilde{m}_{u^c_{B,C}}^2.
\end{align}
This is exactly the soft mass pattern we want as long as $\widetilde{m}_{\overline{P}_2}^2 > \widetilde{m}_{\overline{P}_1}^2$. The different constituent antiquarks may receive the same soft mass because of a symmetry, or from a $U(1)$ gauge mediation where they carry the same charge up to a sign. For example, $\widetilde{m}_{\overline{P}_2}^2$ may be positive because the associated antiquarks have charges $\pm 1$, while all other constituent antiquarks have no charge, so $\widetilde{m}^2_{\overline{P}_1}=\widetilde{m}_{\overline{P}}^2=0$. To give large soft masses to the $A$ sector gluino and light flavor squarks, one can imagine there are also SM colored messenger states which give them large gauge mediated SUSY-breaking masses. The $B/C$ gluinos need not be heavy, so it is not essential to have $B/C$ colored messengers, though they may affect the phenomenology. 

This construction also produces many other composite states (including mesons made of 2 quarks or 2 antiquarks) beyond what are needed in the superpotential. These states can be removed from the low-energy spectrum by marrying them to elementary fields $X_{ij}$ of opposite quantum numbers with superpotential terms like $\kappa X_{ij} \overline{P}_i P_j$. As long as the coupling $\kappa$ is small enough to keep their masses below the confinement scale $\Lambda_G$, they do not affect the confining dynamics. The states which carry SM color must be heavier than a few TeV to avoid experimental constraints.

Assuming the Higgs field is elementary and the top quarks are composite, the top Yukawa couplings arise from higher dimensional superpotential operators above the compositeness scale $\Lambda_G$
\begin{equation}
\sim \frac{g_t}{\Lambda_\text{UV}^2}P\overline{P}P\overline{P}H \,,
\end{equation}
where $\Lambda_\text{UV}$ represents the ultraviolet (UV) cutoff of the theory. They are taken to respect the $Z_3$ symmetry to keep the 3 Yukawa couplings equal. This means that 
\begin{equation}
y_t\sim \frac{g_t\Lambda^2_G}{\Lambda_\text{UV}^2}\,.
\end{equation}
As the top Yukawa is order one, the UV cutoff and the strong scale of the confining group $G$ cannot be very far apart, even though the running due to the strong $G$ gauge coupling can increase $g_t$ at the confinement scale.

To find the constraints on $\Lambda_G$ (and therefore $\Lambda_\text{UV}$), we first notice that Eq.~\eqref{e.mesonMass} is only a leading order result and is corrected by terms of order $m_P^2/\Lambda_G^2$. Consequently, we need $\Lambda_G$ to be significantly larger than the soft masses, i.e., $\Lambda_G \gtrsim 10$ TeV. A similar requirement follows from the masses of the colored composite mesons beyond the top sector. These should be below $\Lambda_G$, but need to be at least a few TeV to satisfy experimental constraints.

These new colored states also contribute to the running of the QCD coupling above their masses. There are 5 additional color-triplets or anti-triplets, 3 made of $\overline{P} P$ as seen in Eq.~\eqref{e.mesonmatrix} and 2 from $PP$-type mesons as the antisymmetric combination of two color-triplets gives an anti-triplet. Together with their elementary partners they contribute 5 flavors beyond the MSSM.
Above their mass (and below $\Lambda_G$) the beta function coefficient for the SM $SU(3)_A$ becomes $b=3N-F=9-11=-2$, so
\begin{equation}
\frac{1}{\alpha_s(\mu)}=\frac{1}{\alpha_s \left(M_X \right)}-\frac{2}{2\pi}\ln\left(\frac{\mu}{M_X} \right),
\end{equation}
where $M_X$ is the threshold scale of these new colored degrees of freedom. Above $\Lambda_G$, there are 4 color-triplet or anti-triplet constituent quarks but one needs to subtract the left-handed top-bottom doublet and the right-handed top. Together with the 5 elementary colored $X$'s this gives $F=(6-3/2)+4/2+5/2=9$ and $b=0$. Although the coupling becomes non-asymptotically free between $M_X$ and $\Lambda_G$, the running is slow and $M_X$ and $\Lambda_G$ do not need to be very far apart, so requiring that the QCD coupling remains finite does not put a strong upper bound on $\Lambda_G$. The $SU(3)_{B,C}$ couplings are also safe, their sectors need not contain the light flavors so they can even be asymptotically free. However, the different particle content in the $A$ and $B, C$ sectors results in different gauge coupling beta functions. Even if the gauge couplings are equal at the cutoff scale, the running induces differences in their values at lower energies. The different gauge couplings feed into the running of the top Yukawa couplings of the $A, B, C$ sectors, affecting the cancelation of the Higgs potential. This is a three-loop effect, so we should have $\Lambda_\text{UV} \lesssim 100$~TeV to avoid large corrections to the Higgs potential. 

In the $A$ sector, the large mass of the gluino affects the soft masses of the stops via renormalization group (RG) running. The leading contribution to the beta function is
\begin{equation} \label{eq:gluinorunning}
\frac{d\,\widetilde{m}^2}{d\ln\mu^2}\approx -\frac{4\alpha_s }{3\pi} \,m_{\tilde{g}}^2\,,
\end{equation}
hence we find a correction
\begin{equation}
\delta \widetilde{m}^2 (m_{\tilde{g}}) \equiv \widetilde{m}^2(m_{\tilde{g}}) - \widetilde{m}^2(\Lambda_G) \approx \frac{8\alpha_s }{3\pi} \,m_{\tilde{g}}^2 \ln \frac{\Lambda_G}{m_{\tilde{g}}}\, ,
\end{equation}
where the running starts from the strong scale $\Lambda_G$ where the composite stops are formed, with masses given by Eq.~\eqref{e.mesonMass}. This, in turn, affects the Higgs potential. In particular, the logarithmically divergent piece now contains an $h^2$ term,
\begin{equation}
\Delta V_{h^2} \approx - \frac{N_c y_t^2}{4\pi^2} h^2 \frac{\delta \widetilde{m}^2 (m_{\tilde{g}})}{2} \ln\frac{\Lambda_G}{\widetilde{m}}= - \frac{N_c y_t^2\alpha_s}{3\pi^3} h^2 m_{\tilde{g}}^2 \ln \frac{\Lambda_G}{m_{\tilde{g}}} \ln\frac{\Lambda_G}{\widetilde{m}}\,,
\end{equation}
where we have approximated the running $\delta \widetilde{m}^2$, which vanishes at $\Lambda_G$, by its ``average" value $\delta \widetilde{m}^2(m_{\tilde{g}})/2$. For $\widetilde{m}\sim m_{\tilde{g}} \sim 2\;\mathrm{TeV}$ and $\Lambda_G\sim10$ TeV the contribution to the Higgs mass parameter is $\Delta m^2 \approx - (160\, \text{GeV})^2$. Notice that the largest subleading terms in the right-hand side of Eq.~\eqref{eq:gluinorunning}, which we have neglected, scale as $\sim +\, y_t^2 \widetilde{m}^2/ (16\pi^2)$ and thus partly reduce the gluino effect.

%%%%%%%%%%%%%%%%%%%%%%%%%%%%%%%%%%%%%%%%%%
\section{Phenomenology\label{sec:pheno}}
This section outlines the relevant experimental bounds on the tripled top framework and highlights some of its distinct phenomenology. We begin by defining the low-energy states. In the $A$ sector, above the top quark (with mass $m_{t_A}^2 = y_t^2 h^2$) we find the stops and left-handed sbottom, whose masses are raised by the soft SUSY breaking to $m^2_{\tilde{t}_A} = m^2_{\tilde{u}_A^c} = \widetilde{m}^2 + y_t^2 h^2$ and $m^2_{\tilde{b}_A} = \widetilde{m}^2$, respectively. Except for some special situations, current LHC searches imply $\widetilde{m}\gtrsim 1\;\mathrm{TeV}$, while the $A$ gluino must have mass larger than about $2$ TeV. Since our construction requires $\widetilde{m} \sim M$, where $M$ is the large SUSY mass, these experimental constraints motivate taking $M \gtrsim 2\;\mathrm{TeV}$, while the other mass scales $\omega$ and $\Delta = \sqrt{M^2 - \widetilde{m}^2}$ are typically much lower, below $1\;\mathrm{TeV}$. 

The $B$ and $C$ sectors have identical spectra, due to the residual $Z_2$ symmetry that relates them. In light of the above considerations about the different mass scales, below $\sim 1\;\mathrm{TeV}$ we expect the following BSM states:
\begin{itemize}
\item {\bf Hidden glueballs}. The absence of light particles charged under the hidden color $SU(3)$ implies that the hadron spectrum is comprised of glueballs. There are several of these states, and the lightest is known \cite{Morningstar:1999rf,Chen:2005mg,Meyer:2008tr} to have $J^{PC} = 0^{++}$ and mass related to the confinement scale by $m_0\approx 6.8\, \Lambda_{\text{QCD}_{B,C}}$. 
\item {\bf Top siblings and cousins}. These are the electroweak singlet and doublet $Z_3$ copies of the top sector. The singlet states $\tilde{u}^c_{B,C}$, which are scalars with mass set by $\Delta$, are responsible for canceling the quadratic divergence in the Higgs mass coming from the top loop. Taking our cues from the many proposals to enlarge the Higgs family, we refer to these hidden copies of the stop as top ``siblings." The doublet states, which include both fermion and scalar components of $Q_{B,C}$ and $Q^{\prime c}_{B,C}$, have instead masses set by $\omega$. These particles are also related to the top, but not as closely, so we refer to them as top ``cousins." When the Higgs gets a VEV the siblings and the scalar cousins mix, but we still refer to them by the family title that dominates, so a mostly-$\Delta$ state is called a sibling and a mostly-$\omega$ state is called a cousin. 
\end{itemize}
Because the siblings with mass $\sim \Delta$ are mostly-SM-singlet states, they are difficult to probe experimentally. They do affect the Higgs self coupling and wave function renormalization, but even at proposed future colliders the reach on their mass through these indirect probes is very limited~\cite{Craig:2013xia,Curtin:2015bka}. Since the fine tuning required to make $\Delta$ small scales as $\Delta^2/M^2$, we restrict our analysis to $\Delta > 100\;\mathrm{GeV}$, neglecting the region where the tuning becomes extreme.\footnote{The region with $\Delta < 100\;\mathrm{GeV}$ is also subject to important constraints from $Z$ and Higgs decays.} Then, as shown below, the main signals originate from the cousins, which carry SM electroweak charges and have masses set by $\omega$.

\begin{figure}[t]
\includegraphics[width=10cm]{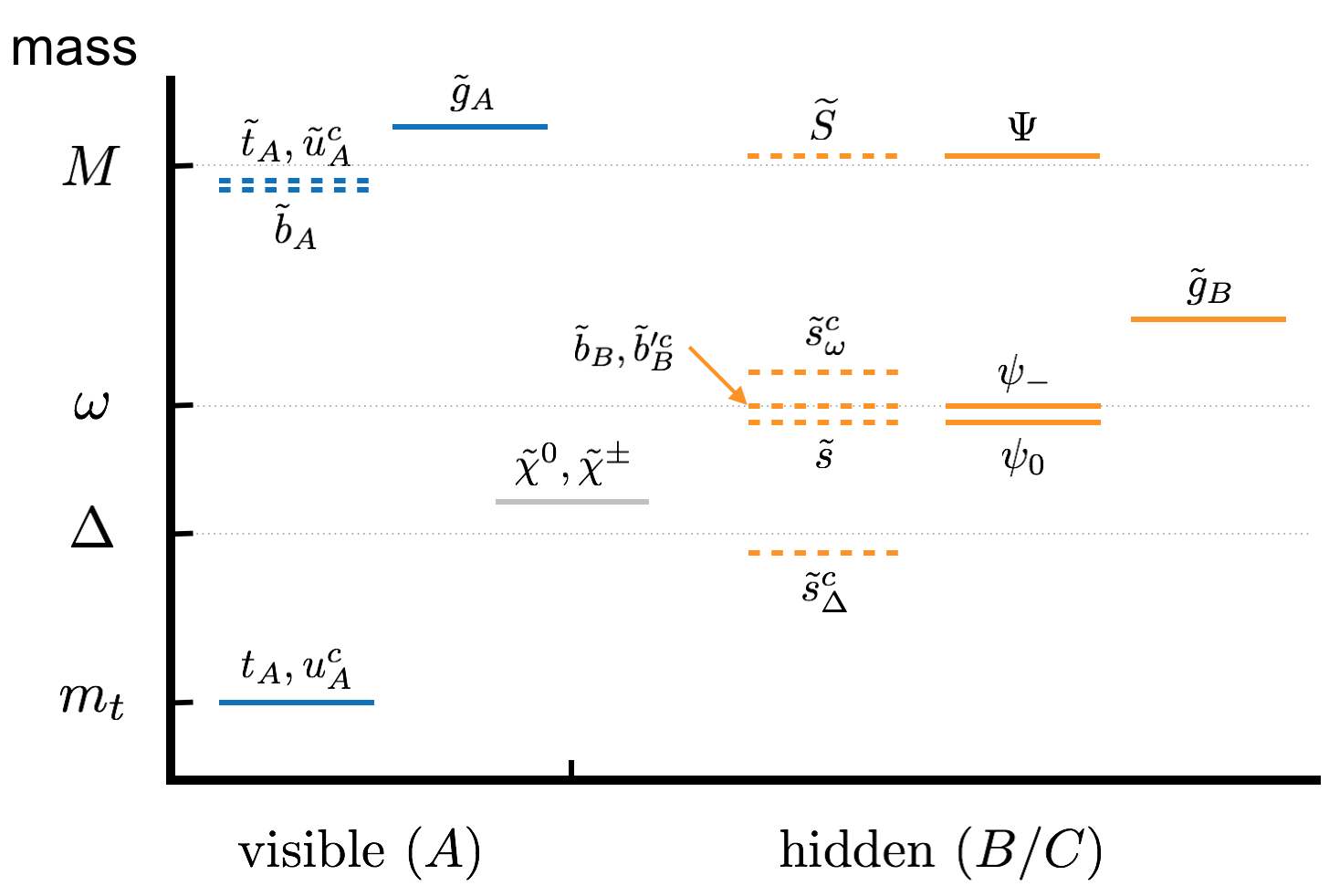}
\caption{Illustrative tripled top spectrum with $\Delta < \omega$. We show only the $B$ sector states as the $C$ sector is identical, due to the residual $Z_2$ symmetry.}
\label{f.tripletStops}
\end{figure}

While light hidden color glueballs and top siblings and cousins are essential ingredients of the model, some other SUSY particles may also be below $1\;\mathrm{TeV}$ as their experimental bounds are weaker than those of SM colored particles: 
\begin{itemize}
\item {\bf MSSM sleptons, charginos and neutralinos} (collectively denoted as electroweakinos or EWinos). These are much less constrained than the squarks and gluino, and may have masses much below $1$ TeV. 
%\
\item {\bf Gluinos of the $B$ and $C$ sectors}.
\end{itemize}
The possible spectra of these particles, see for example Fig.~\ref{f.tripletStops}, lead to a rich and varied phenomenology. In this paper we focus on a few representative scenarios, leaving a more complete analysis for future study.

We first determine the mass eigenstates of the hidden sectors. The mass matrix for the $B$ sector fermions is
\begin{equation}
- \begin{pmatrix} u^\prime_B & t_B \end{pmatrix} \mathcal{M}_F \begin{pmatrix} u_B^c \\ t_B^{\prime c} \end{pmatrix},  \qquad \mathcal{M}_F =  \begin{pmatrix} M & 0 \\ y_t h & \omega \end{pmatrix},
\label{e.fermionMassMat}
\end{equation}
which is diagonalized by $R(\theta_L)^T \mathcal{M}_F R(\theta_R) = \mathrm{diag}\,(M_\Psi, m_{\psi_0})$, where the physical masses are, assuming $M>\omega$ and $M \gg  y_t  h \,$,
\begin{equation}
M_{\Psi}^2\approx M^2\left(1+\frac{y_t^2h^2}{M^2-\omega^2} \right), \qquad m_{\psi_0}^2\approx\omega^2\left(1-\frac{y_t^2h^2}{M^2-\omega^2} \right) .
\end{equation}
The rotations read
\begin{equation}
\begin{pmatrix} u^\prime_B \\ t_B \end{pmatrix} \to R(\theta_L) \begin{pmatrix} \Psi \\ \psi \end{pmatrix}, \quad \begin{pmatrix} u_B^c \\ t_B^{\prime c} \end{pmatrix} \to R(\theta_R) \begin{pmatrix} \Psi^c \\ \psi^c \end{pmatrix}, \qquad R(\theta) \equiv \begin{pmatrix} \cos\theta & \sin\theta \\ - \sin \theta & \cos\theta \end{pmatrix}, 
\end{equation}
with mixing angles given by
\begin{equation}
\sin \theta_L \approx - \frac{y_t h M}{M^2-\omega^2}\,, \qquad \sin \theta_R \approx - \frac{y_t h \omega}{M^2-\omega^2}\,.
\end{equation}
Hence $\psi, \psi^c$ form an electrically neutral Dirac fermion $\psi_0$, whereas the $SU(2)_L$ partner states $b_B, b_B^{\prime c \,\dagger}$ form a Dirac fermion $\psi_-$ with electric charge $-1$ and mass $m_{\psi_-}=\omega$. 

The $B$ sector scalar masses are given by
\begin{equation} \label{e.scalarMassMat}
- \begin{pmatrix} \tilde{u}^\prime_B & \tilde{t}_B \end{pmatrix}^\ast \mathcal{M}^2_S \begin{pmatrix} \tilde{u}^\prime_B \\ \tilde{t}_B \end{pmatrix},\qquad  \mathcal{M}^2_S = \begin{pmatrix} M^2 & y_t h M \\ y_t h M & \omega^2 + y_t^2 h^2 \end{pmatrix} ,
\end{equation}
\begin{equation}\label{e.scalarcMassMat}
- \begin{pmatrix} \tilde{u}_B^c  & \tilde{t}^{\,\prime c}_B  \end{pmatrix}  \mathcal{M}^2_{S^c} \begin{pmatrix} \tilde{u}_B^c   \\ \tilde{t}_B^{\,\prime c} \end{pmatrix}^\ast ,\qquad  \mathcal{M}^2_{S^c} = \begin{pmatrix} \Delta^2 + y_t^2 h^2  & y_t h \omega \\ y_t h \omega & \omega^2 \end{pmatrix} ,
\end{equation}
where $\Delta^2 = M^2 - \widetilde{m}^2$. The $\mathcal{M}^2_S$ matrix is not affected by SUSY breaking, hence it is diagonalized by a rotation $R(\phi_L)$ with $\phi_L = \theta_L$, yielding a heavy mass eigenstate $\widetilde{S}$ with $M^2_{\widetilde{S}} = M^2_\Psi \sim M^2$, and a light mass eigenstate $\tilde{s}$ with $m^2_{\tilde{s}} = m^2_{\psi_0} \sim \omega^2$. The $\mathcal{M}^2_{S^c}$ matrix requires special attention. While the other particle mixings are suppressed by $M$ and are therefore small, in this case the large negative soft mass $ - \widetilde{m}^2 |\tilde{u}_B^c|^2$ causes more uniform mixing. Diagonalization is achieved through the rotation
\begin{equation} \label{eq:M2Sc}
\begin{pmatrix} \tilde{u}_B^c   \\ \tilde{t}_B^{\,\prime c} \end{pmatrix} \to R(\phi_R) \begin{pmatrix} \tilde{s}^c_\Delta  \\ \tilde{s}^c_\omega  \end{pmatrix},\qquad \sin 2\phi _{R}= \frac{2y_t h\omega}{ m_2^2- m_1^2}\, \mathrm{sgn} \left(\omega^2 - \Delta^2 - y_t^2 h^2\right)\,,
\end{equation}
where the mass eigenvalues are
\begin{equation} \label{eq:siblingcousinmasses}
m_{2,1}^2=\frac{1}{2} \left(\omega^2+\Delta^2+y_t^2h^2\pm\sqrt{\left( \omega^2+\Delta^2 + y_t^2 h^2\right)^2 - 4 \omega^2 \Delta^2}\, \right),
\end{equation}
resulting in $R(\phi_R)^T \mathcal{M}^2_{S^c} R(\phi_R) = \mathrm{diag}\,(m^2_{\tilde{s}^c_\Delta} \sim \Delta^2,  m^2_{\tilde{s}^c_\omega} \sim \omega^2)$, i.e., $\tilde{s}^c_\Delta$ is a sibling, and $\tilde{s}^c_\omega$ is a cousin.\footnote{Note that for $\Delta^2 + y_t^2 h^2 < \omega^2 \,(\Delta^2 + y_t^2 h^2 >\omega^2)$ we have $m_{\tilde{s}^c_\Delta} = m_{1}\,(m_{\tilde{s}^c_\Delta} = m_2)$, i.e., the sibling is lighter (heavier) than the cousin. When $ \Delta^2+y_t^2 h^2 = \omega^2$ the mixing is maximal.} Finally, the scalars $\tilde{b}_B, \tilde{b}_B^{\prime c}$ have electric charges equal to $-1$ and $+1$, respectively, and masses equal to $\omega$. 

From Eq.~\eqref{eq:siblingcousinmasses}, the couplings to the Higgs of the heavier and lighter neutral scalar are
\begin{equation}
\frac{y_t^2}{2}\left(1\pm\frac{\omega^2+\Delta^2}{\left|\omega^2-\Delta^2 \right|} \right),
\end{equation}
respectively (notice that in either limit of $\Delta\to 0$ or $\omega\to 0$ the coupling to the lighter state completely vanishes). The sum over the two states yields the usual coupling $y_t^2$ for one stop, and this is why two copies of this structure, i.e., both the $B$ and $C$ sectors, are needed to completely cancel the top loop.

We note that the scalar mass matrices in Eqs.~\eqref{e.scalarMassMat} and~\eqref{e.scalarcMassMat} only include the leading SUSY-breaking effects. In general, additional SUSY-breaking terms should be considered, including extra diagonal soft masses for all the fields, as well as A- and B-terms, 
\begin{equation} \label{eq:extraAandBterms}
V_{\rm s}\; \ni \;   y_t A_t \widetilde{Q}_B H \tilde{u}_B^c + B_Q \widetilde{Q}_B \widetilde{Q}_B^{\prime c} + B_u \tilde{u}_B^{\prime} \tilde{u}^c_B + \mathrm{h.c.} 
\end{equation}
(if these mixing terms are present, $\mathcal{M}^2_S$ and $\mathcal{M}^2_{S^c}$ are combined into a single $4\times 4$ mass matrix). These extra soft terms are radiatively generated and can be much smaller than $\omega$ and $\Delta$, thus giving only suppressed corrections to the physical scalar masses.

For simplicity, throughout our discussion we assume that the Higgs sector is in the decoupling limit at large $\tan\beta$, which implies the VEV satisfies $\langle h\rangle \approx v/\sqrt{2}$.

\subsection{Hidden glue dynamics\label{ssec.HidGlueballs}} 
\begin{figure}[t]
 \begin{center}
\includegraphics[width=0.55\textwidth]{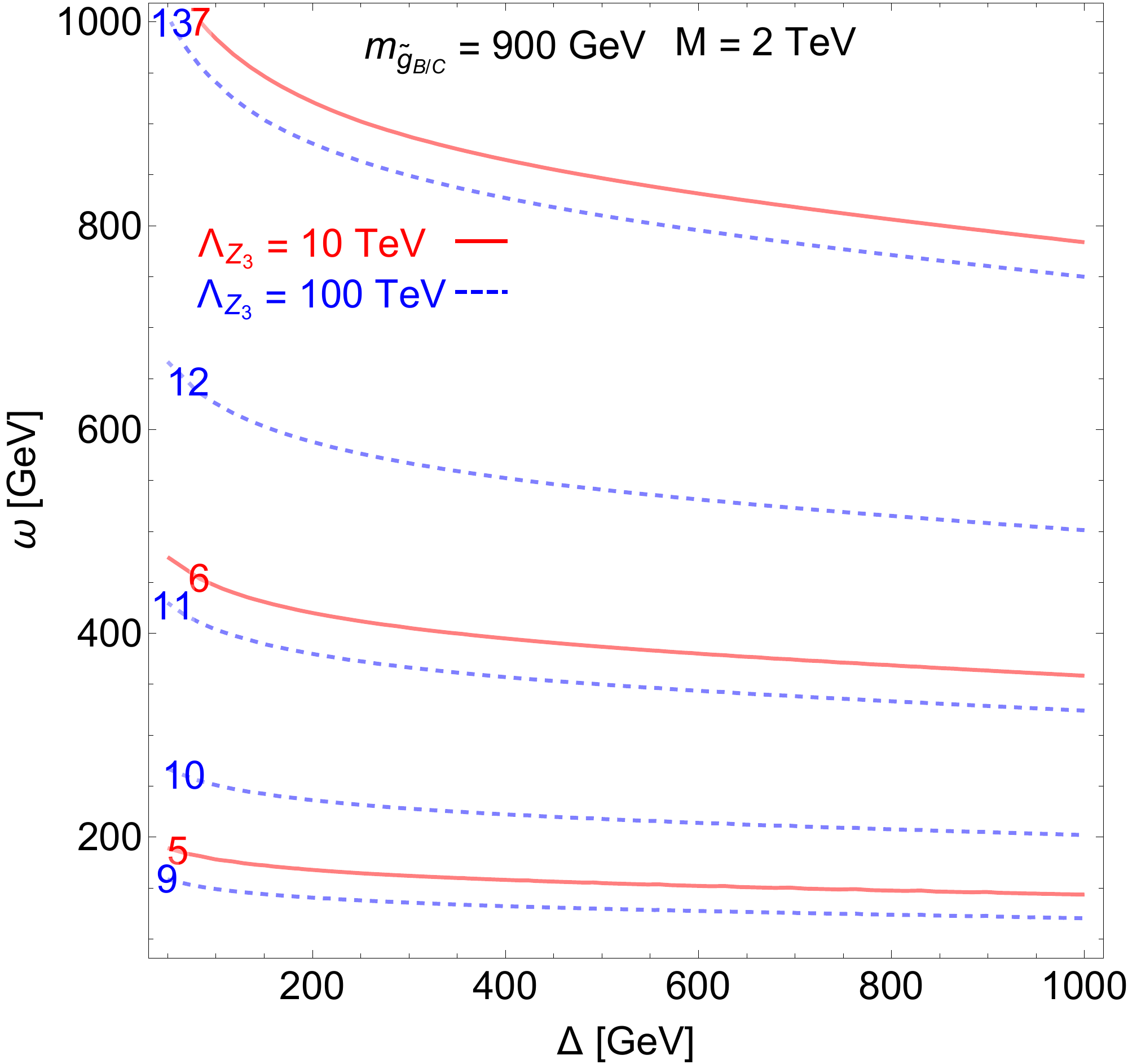} 
 \end{center}
 \caption{Contours of $\Lambda_{\text{QCD}_{B,C}} / \mathrm{GeV}$ in the $(\Delta, \omega)$ plane. The large SUSY mass is taken to be $M = 2\;\mathrm{TeV}$ and the hidden gluino mass $900\;\mathrm{GeV}$. Solid red (dashed blue) contours correspond to a scale of unbroken $Z_3$ of $10\,(100)\,\mathrm{TeV}$. The mass of the lightest glueball is $m_0\approx 6.8\,\Lambda_{\text{QCD}_{B,C}}$.}
\label{f.lambdaQCD}
\end{figure}

The mass of the lightest hidden glueball is related to the confinement scale by $m_0\approx 6.8\,\Lambda_{\text{QCD}_{B,C}}$. The confinement scale can be computed using RG running, as a function of the scale at which the strong couplings in the three sectors are assumed to be equal, $\Lambda_{Z_3}$, and of the particle masses. We take $\Lambda_{Z_3}$ to be in the range of $10$--$100\;\mathrm{TeV}$, as suggested by the cutoff scale $\Lambda_{\rm UV}$ of the model discussed at the end of Sec.~\ref{s.construct}. Performing the renormalization group (RG) running at two loops (see Appendix \ref{a.lambdaQCD} for details) we find a range of $\Lambda_{\text{QCD}_{B,C}}$ values from about 2.5 to 16 GeV. In Fig.~\ref{f.lambdaQCD} we show the dependence of the confinement scale on the hidden sector masses: In most of the parameter space we focus on, we have $ \Lambda_{\text{QCD}_{B,C}} \gtrsim 4\;\mathrm{GeV} $, which corresponds to $  m_0 \gtrsim 28\;\mathrm{GeV}$. All other glueball masses are known from the lattice~\cite{Morningstar:1999rf,Chen:2005mg,Meyer:2008tr} in terms of $m_0$. 

Since the lightest hidden glueball has $0^{++}$ quantum numbers, it can mix with the Higgs boson~\cite{Juknevich:2009gg} through the effective coupling
\begin{equation} 
\frac{c_g \alpha_{d}}{12\pi}\frac{h}{v} \, \hat{G}^{\,a}_{\mu\nu} \hat{G}^{\,a\,\mu\nu} \,, \label{e.HiggsGlue}
\end{equation}
where $\alpha_{d} = g_{d}^2/ (4\pi)$ with $g_{d}$ denoting the strong coupling of the $B$ or $C$ sector, and $\hat{G}^{\,a\,\mu\nu}$ is the corresponding gluon field strength. The coupling $c_g$ is induced by loops of particles that are charged under hidden color and couple to the Higgs, i.e., the top siblings and cousins. The operator in Eq.~(\ref{e.HiggsGlue}) allows the lightest glueball to decay to SM particles through an off-shell Higgs. Denoting a given SM state with $Y$, the decay width is
\begin{equation}
\Gamma(0^{++} \to YY) = |c_g|^2 \left( \frac{\alpha_d}{6\pi} \frac{f_{0^{++}}}{v(m_h^2 - m_0^2)} \right)^2 \Gamma (h (m_0) \to YY)_{\rm SM}\,.
\end{equation}
The decay constant is defined by $ f_{0^{++}} = \langle 0 | \mathrm{Tr} \,\hat{G}_{\mu\nu} \hat{G}^{\mu \nu} | 0^{++} \rangle$, whereas $\Gamma (h (m_0) \to YY)_{\rm SM}$ is the partial width of a SM Higgs with mass $m_0$. Lattice results in pure-glue $SU(3)$ give $4\pi\alpha_d f_{0^{++}} = 3.1 m_0^3 $ \cite{Chen:2005mg}.

We can estimate the size of $c_g$ using the Higgs low-energy theorem (LET)~\cite{Ellis:1975ap,Shifman:1979eb} when all the particles that mediate it are heavier than $\sqrt{p_h^2}/2$, with $p_h$ the Higgs four-momentum. Since in the \mbox{$0^{++} \to YY$} decay we have $\sqrt{p_h^2}/2 = m_0/2 \lesssim 50\;\mathrm{GeV}$, the LET applies throughout our parameter space. Treating the Higgs as a background field and viewing the field-dependent mass $M_i (h)$ of each heavy particle as a threshold for the running of $g_d$, we write the low-energy Lagrangian
\begin{equation}
\mathcal{L}_\text{LET}=\frac{\alpha_d}{16\pi}\hat{G}^a_{\mu\nu}\hat{G}^{a\mu\nu}\sum_i  \delta b_i \ln M^2_i(h),
\end{equation}
where the beta function coefficient is $\delta b=2/3\,(1/6)$ for a Dirac fermion (complex scalar) and we have assumed that the virtual particles transform in the fundamental representation of $SU(3)$. Expanding to first order in $h$ we arrive at the coupling in Eq.~\eqref{e.HiggsGlue} with 
\begin{equation} \label{eq:cgLET}
c_g^{\rm LET} = \, v \left[\frac{\partial}{\partial h}\ln\det\mathcal{M}_f(h)+\frac{1}{8}\frac{\partial}{\partial h}\ln \det\mathcal{M}^2_s (h) \right]_{\langle h\rangle},
\end{equation}
where $\mathcal{M}_f$ and $\mathcal{M}^2_s$ are the fermion and scalar mass matrix in the Higgs background, respectively.\footnote{Notice that Eq.~\eqref{eq:cgLET} assumes canonical normalization for the background field.} The mass matrices in Eqs. \eqref{e.fermionMassMat}, \eqref{e.scalarMassMat}, and \eqref{e.scalarcMassMat} all have $h$-independent determinants, hence Eq.~\eqref{eq:cgLET} vanishes. Corrections to this leading-order result arise from subleading SUSY-breaking terms in the scalar mass matrices, as well as from subleading terms in the expansion of the form factors. We find that the extra soft terms give the most important effects, which we estimate by adding a universal contribution $\delta m^2$ to the scalar masses. Then the largest correction comes from $\mathcal{M}^2_{S^c}$, which to leading order in $\delta m^2 \ll \omega^2,\Delta^2 $ yields
\begin{equation} \label{eq:cgLETcorrection}
c_g^\text{LET} \simeq \frac{\delta m^2 m_t^2}{4\omega^2\Delta^2}\,.
\end{equation}
From this result we estimate the glueball's proper decay length
\begin{equation} 
c\tau_{0^{++}} \sim 1.2\,\text{m}\left(\frac{5\;\text{GeV}}{\Lambda_{\text{QCD}_{B,C}}} \right)^7\left(\frac{\omega}{500\,\text{GeV}} \right)^4\left(\frac{\Delta}{300\,\text{GeV}}  \right)^4\left(\frac{100\,\text{GeV}}{\delta m}   \right)^4, \label{e.glueballctau}
\end{equation}
where we have used the benchmark $5\;\mathrm{GeV}$ for the hidden confinement scale, typical for $\Lambda_{Z_3} \sim 10\;\mathrm{TeV}$. For comparison, in FSUSY the glueball decay length is a few millimeters for similar values of the parameters, so in our model the hidden glueballs are relatively long-lived. We stress, however, that this estimate of the glueball lifetime has large uncertainties due to the high-power dependences on the model parameters. In Eq.~\eqref{e.glueballctau}, a mild suppression of $\delta m$ can easily push the decay length beyond $10$ meters, making the glueballs decay mostly out of the LHC detectors. Conversely, for a larger confinement scale $\Lambda_{\text{QCD}_{B,C}} \sim 10\;\mathrm{GeV}$ (corresponding to a higher $\Lambda_{Z_3} \sim 100\;\mathrm{TeV}$) a moderate enhancement of $\delta m$ can lead to a sub-millimeter lifetime. This makes the identification of the $0^{++} \to b\bar{b}$ displaced vertices challenging, although this may improve in the near future~\cite{Ito:2018asa}. 

If $m_0$ is smaller than $m_h/2$, the Higgs has exotic decays into pairs of $0^{++}$ hidden glueballs, a signature that has been carefully analyzed in the context of Neutral Naturalness~\cite{Craig:2015pha,Curtin:2015fna,Csaki:2015fba}. The rate is again controlled by the expression of $c_g$ in Eq.~\eqref{eq:cgLETcorrection}. The width for decay to the gluons of one sector reads
\begin{equation}
\Gamma(h\to g_{B}g_{B}) = \frac{\alpha_d (m_h/2)^2 m_h^3}{72 \pi^3 v^2}\left|c_g \right|^2, 
\end{equation}
yielding a branching ratio
\begin{equation}
\text{BR}\left(h\to g_B g_B + g_Cg_C \right) \sim 2\cdot 10^{-6}\left( \frac{\alpha_d(m_h/2)}{0.17}\right)^2\left( \frac{\delta m}{100\,\text{GeV}}\right)^4\left(\frac{500 \,\text{GeV}}{\omega} \right)^4\left( \frac{300\,\text{GeV}}{\Delta}\right)^4, 
\end{equation}
which is suppressed compared to FSUSY. The smaller branching ratio makes detection of these exotic Higgs decays at the LHC extremely challenging, but they may be within reach of a future $100$ TeV collider, either with the main detectors or with MATHUSLA~\cite{Chou:2016lxi,Curtin:2017izq}, depending on the glueball lifetime.

\subsection{Quirky signals from cousins\label{ssec.QuirkSig}} 
The cousin particles are composed of both fermions and scalars, all carrying electroweak charges and with masses around $\omega$. Because several among them have unit electric charge, LEP2 direct constraints on charged particles (see Ref.~\cite{Egana-Ugrinovic:2018roi} for a recent appraisal) imply $\omega\gtrsim 100\;\mathrm{GeV}$. Fermions have a larger production cross section in quark-antiquark annihilation, hence we discuss the signals from fermionic cousins first, assuming that their decays to the sibling and cousin scalars are kinematically forbidden (e.g., $\Delta > \omega$ and $\delta m^2>0$). 
The discussion of the signals from the cousin and sibling scalars is presented in Secs.~\ref{ssec.SquirkSigCousin} and~\ref{ssec.SquirkSigSibling}, respectively. 

When a cousin fermion-antifermion pair is produced through the Drell-Yan (DY) process, it is connected by an $SU(3)_{B,C}$ color flux string. Since the hidden sector has no light matter particles that can be pair produced to break the string, the pair remains tied, forming a highly excited ``quirky'' bound state (quirkonium)~\cite{Okun:1980mu,Okun:1980kw,Kang:2008ea}. Annihilation from bound states with $\ell > 0$ is highly suppressed~\cite{Kang:2008ea}, so the system must de-excite down to one of the two lowest-lying $s$-wave states, the spin-0 $\eta$ or spin-1 $\Upsilon$, before it can efficiently annihilate.

If at least one of the quirks is electrically charged, the system can de-excite by emission of soft photons \cite{Burdman:2008ek,Harnik:2008ax}. Consider a pair of quirks of mass $m_\psi$ with kinetic energy $K$, connected by a string with tension $\sigma\approx 3.6\,\Lambda_{\text{QCD}_{B,C}}^2$~\cite{Lucini:2004my}. The acceleration of the quirks due to the constant force exerted by the string is $a = \sigma/m_\psi$. The power $\mathcal{P}$ radiated by the accelerating charges is given by the Larmor formula,
\begin{equation}
\mathcal{P}=\frac{8\pi\alpha}{3}\,a^2 = \frac{8\pi\alpha}{3}\frac{\sigma^2}{m_\psi^2}\,.\label{e.larmor}
\end{equation}
The de-excitation time is obtained dividing the kinetic energy by the power,
\begin{equation}
t^{\,\gamma}_{\,\text{de-excite}}\sim\frac{K}{\mathcal{P}}=\frac{3 m_\psi^2K}{8\pi \alpha\sigma^2}\,.\label{e.photonTimeGen}
\end{equation}
For a typical initial kinetic energy $K \sim m_\psi$, we find
\begin{equation}
t^{\,\gamma}_{\,\text{de-excite}}\sim  2 \cdot 10^{-19}\;\mathrm{s}\, \left( \frac{\omega}{500\;\mathrm{GeV}} \right)^3 \Big(\frac{5\;\mathrm{GeV}}{ \Lambda_{\text{QCD}_{B,C}}} \Big)^4\,,
\label{e.photonTime}
\end{equation}
where we have used $m_\psi \sim \omega$. The charged quirk can also beta decay to the neutral one, with width given by
\begin{equation} \label{eq:betadecayfermions}
\Gamma_\beta \simeq \frac{3G_F^2 (\Delta m_\psi)^5}{5\pi^3}\,.
\end{equation}
We have defined the mass splitting of charged and neutral quirks as
\begin{equation} \label{e.quirksplitting}
\Delta m_\psi = m_{\psi_-} - m_{\psi_0} \simeq \frac{m_t^2}{2(M^2 - \omega^2)}\,\omega\,,
\end{equation}
which is typically a few GeV: for example, for $M = 2\;\mathrm{TeV}$ and $\omega = 500\;\mathrm{GeV}$ we find $\Delta m_\psi \simeq 2.0\;\mathrm{GeV}$. For our typical parameters the width in Eq.~\eqref{eq:betadecayfermions} corresponds to $t_\beta \sim 10^{-14}\;\mathrm{s}$, which is much longer than $t^{\,\gamma}_{\,\text{de-excite}}$. Thus, soft photon emission enables $\overline{\psi}_+ \psi_-$ and $\overline{\psi}_0 \psi_-$ pairs to de-excite promptly to the ground state. 

The situation is different for the $\overline{\psi}_0 \psi_0$ pair, which does not couple directly to the photon. In this case hidden gluons can still be radiated, but these cannot be softer than the mass of the lightest glueball, which may lead to a kinematic suppression as large as $(\sqrt{\sigma}/m_0)^6\sim5\, \cdot \,10^{-4}$~\cite{Kang:2008ea}. Conversely, the hidden gluons couple much more strongly than photons, $\alpha_d/\alpha\sim 25$. The resulting timescale is $t^{\,\rm glueball}_\text{\,de-excite} \lesssim 10^{-17}\;\mathrm{s}$, which is still prompt. However, hidden glueball radiation cannot completely de-excite the system. At first the emission of glueballs proceeds rapidly, but as soon as the quirks reach a kinetic energy $K\lesssim m_0$, it becomes kinematically forbidden. For a linear potential $V(r) = \sigma r$, the energy levels are approximately given by
\begin{equation}
E_n\approx \left(\frac{3\pi}{2} \right)^{2/3}\frac{\sigma^{2/3}}{(2\mu)^{1/3}}\left(n-\frac14 \right)^{2/3},\label{e.energystates}
\end{equation}
where $n\geq 1$ and $\mu = m_\psi/2$ is the reduced mass. Hence, a quirk pair with kinetic energy $K\lesssim m_0$ can have $n$ as high as
\begin{equation} 
n \sim \frac{2 }{3\pi} \frac{m_\psi^{1/2} m_0^{3/2}}{\sigma} + \frac{1}{4} \approx 10 \left( \frac{\omega}{500\;\mathrm{GeV}} \right)^{1/2} \left(\frac{5\;\mathrm{GeV}}{ \Lambda_{\text{QCD}_{B,C}}} \right)^{1/2}\,.\label{e.nestimate}
\end{equation}
Notice that for such $n$ the potential is safely dominated by its linear component. Equation~\eqref{e.nestimate} shows that glueball radiation alone is unlikely to reach the system's ground state. 

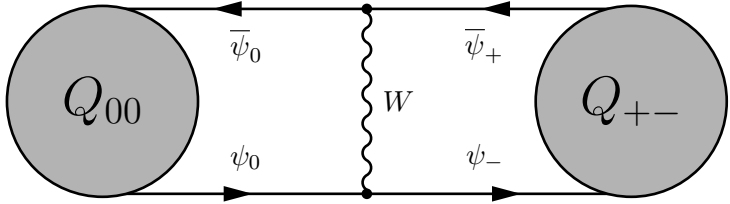
\begin{figure}[t]
\begin{fmffile}{MesonMix}
\begin{fmfgraph*}(200,70)
\fmfpen{1.0}
\fmfstraight
\fmfleft{i1,m1,i2}\fmfright{o1,m2,o2}
\fmf{fermion,tension=1.0}{i1,v1,o1}
\fmf{fermion,tension=1.0}{o2,v2,i2}
\fmffreeze
\fmfv{l=$\psi_{0}$,l.d=50,l.a=16}{i1}
\fmfv{l=$\overline{\psi}_{0}$,l.d=50,l.a=-16}{i2}
\fmfv{l=$\psi_{-}$,l.d=50,l.a=164}{o1}
\fmfv{l=$\overline{\psi}_{+}$,l.d=50,l.a=-164}{o2}
\fmf{boson,tension=0.5,label=$W$,label.side=right}{v1,v2}
\fmfv{decor.shape=circle,decor.filled=full,decor.size=1.5thick}{v1} 
\fmfv{decor.shape=circle,decor.filled=full,decor.size=1.5thick}{v2} 
\fmfv{decor.shape=circle,decor.filled=30,decor.size=1in,l=\LARGE$Q_{00}$,l.a=0,l.d=0}{m1}
\fmfv{decor.shape=circle,decor.filled=30,decor.size=1in,l=\LARGE$Q_{+-}$,l.a=0,l.d=0.0}{m2}
\end{fmfgraph*}
\end{fmffile}
\caption{\label{f.MesonMix} Mass mixing between the quirk bound states $Q_{00} = \overline{\psi}_0 \psi_0$ and $Q_{+-} = \overline{\psi}_+ \psi_-\,$, induced by the exchange of a virtual $W$ boson.}
\end{figure}

The $Q_{00} \equiv \overline{\psi}_{0} \psi_{0}$ bound state does not couple to photons directly, but it is also not a mass eigenstate. It mixes with $Q_{+-} \equiv \overline{\psi}_+ \psi_-$ through the $t$-channel exchange of a $W$ boson, as shown in Fig.~\ref{f.MesonMix}. After this mixing is diagonalized, the resulting mostly-$Q_{00}$ eigenstate inherits a non-vanishing decay width to photons. The corresponding lifetime is estimated as (see Appendix~\ref{a.QuirkoniumMixing} for the calculation) 
\begin{equation}
t_{\gamma}^{00} \sim \frac{(\Delta m_\psi)^2}{2 G_F^2 \,\omega^2 \Lambda_{\text{QCD}_{B,C}}^4 \Gamma_\gamma}  \sim 10^{-17}\;\mathrm{s}\, \bigg( \frac{5\;\mathrm{GeV}}{\Lambda_{\text{QCD}_{B,C}}} \bigg)^8 \bigg(\frac{2\;\mathrm{TeV}}{M} \bigg)^4 \bigg( \frac{\omega}{500\;\mathrm{GeV}} \bigg)^3\,,
\label{e.mixingGamma}
\end{equation}
where $G_F = 1/(\sqrt{2} v^2)$ is the Fermi constant, and $\Gamma_\gamma$ is the width of $Q_{+-}$ for photon emission, related to the timescale in Eq.~\eqref{e.photonTime} by $\Gamma_\gamma = 1/t^{\,\gamma}_{\,\text{de-excite}}$. The result in Eq.~\eqref{e.mixingGamma} shows that the mixing between the $Q_{00}$ and $Q_{+-}$ mesons is large enough for the former to de-excite promptly to the ground state via soft photon emission. Glueball radiation may accelerate the de-excitation process.
 
Having established that all quirk pairs promptly de-excite via photon radiation to the ground state and annihilate, we analyze the resulting resonant signals at the LHC. We discuss first the electrically charged channel, whose DY cross section is larger than those of the neutral channels and electric charge conservation forbids decays to hidden gluons only, leading to increased branching fractions to SM particles. The partonic cross section for open production is (defining $\alpha_W = g^2/(4\pi)$) 
\begin{equation} \label{eq:DYviaWexchange}
\hat{\sigma}(u\bar{d} \to \overline{\psi}_+ \psi_0) \simeq \frac{\pi \alpha_W^2}{6\,\hat{s}} \frac{\hat{s}^2}{(\hat{s} - m_W^2)^2} \left(1 - \frac{4 \omega^2}{\hat{s}} \right)^{1/2} \bigg( 1 + \frac{2 \omega^2}{\hat{s}} \bigg) ,
\end{equation}
from which we obtain the hadronic cross section by convoluting with the parton luminosities.\footnote{We use MSTW2008 NLO parton distribution functions \cite{Martin:2009iq} with factorization scale set to $\sqrt{\hat{s}}/2$.} For example, setting $\omega = 500\;\mathrm{GeV}$ we find $\sigma(pp \to  \overline{\psi}_+ \psi_{0} + \overline{\psi}_{0} \psi_-) = 59\;\mathrm{fb}$ at $13\;\mathrm{TeV}$. In our analysis of quirkonium production and decay we neglect the small mixing between $\psi_0$ and the heavy fermion $\Psi$, which is suppressed by the large scale $M$. After de-excitation the quirk pair annihilates, giving resonant signals of invariant mass $\sim 2\omega$. 

\begin{figure}[t]
\begin{subfigure}[b]{0.49\textwidth}
\begin{fmffile}{etaDecay}
\begin{fmfgraph*}(150,70)
\fmfpen{1.0}
\fmfstraight
\fmfleft{i1,m1,i2}\fmfright{o1,m2,o2}
\fmf{fermion,tension=1.0}{i1,v1}\fmf{boson,tension=1.4}{v1,o1}
\fmf{fermion,tension=1.0}{v2,i2}\fmf{boson,tension=1.4}{v2,o2}
\fmfv{l=$\psi_{0}$,l.d=50,l.a=16}{i1}
\fmfv{l=$\overline{\psi}_{+}$,l.d=50,l.a=-16}{i2}
\fmfv{l=$Z,,\gamma$,l.d=5,l.a=0}{o2}
\fmfv{l=$W^{+}$,l.d=5,l.a=0}{o1}
\fmf{fermion,tension=0.3,label=$\psi_{-}$,label.side=right}{v1,v2}
\fmfv{decor.shape=circle,decor.filled=full,decor.size=1.5thick}{v1} 
\fmfv{decor.shape=circle,decor.filled=full,decor.size=1.5thick}{v2} 
\fmfv{decor.shape=circle,decor.filled=30,decor.size=1in,l=\LARGE$\eta_{+0}$,l.a=0,l.d=0}{m1}
\end{fmfgraph*}
\end{fmffile}
\end{subfigure}
\begin{subfigure}[b]{0.49\textwidth}
\begin{fmffile}{upsilonDecay}
\begin{fmfgraph*}(150,70)
\fmfpen{1.0}
\fmfstraight
\fmfleft{i1,m1,i2}\fmfright{o1,m2,o2}
\fmf{fermion,tension=0.4}{i1,v1,i2}
\fmf{fermion,tension=1.5}{o1,v2,o2}
\fmfv{l=$\psi_{0}$,l.d=50,l.a=12}{i1}
\fmfv{l=$\overline{\psi}_{+}$,l.d=50,l.a=-12}{i2}
\fmfv{l=$\ell^{+}$,l.d=5,l.a=0}{o2}
\fmfv{l=$\nu$,l.d=5,l.a=0}{o1}
\fmf{boson,tension=1.5,label=$W^{+}$,label.side=right}{v1,v2}
\fmfv{decor.shape=circle,decor.filled=full,decor.size=1.5thick}{v1} 
\fmfv{decor.shape=circle,decor.filled=full,decor.size=1.5thick}{v2} 
\fmfv{decor.shape=circle,decor.filled=30,decor.size=1.0in,l=\LARGE$\Upsilon_{+0}$,l.a=0,l.d=0}{m1}
\end{fmfgraph*}
\end{fmffile}
\end{subfigure}
\caption{\label{f.QuirkDecays} Dominant decays of the $\eta_{+0}$ and $\Upsilon_{+0}\,$.}
\end{figure}
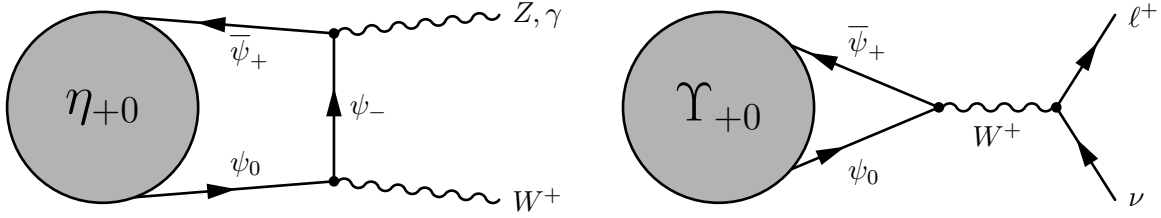

The phenomenology of quirky bound states was studied in several scenarios~\cite{Cheung:2008ke,Kribs:2009fy,Martin:2010kk,Harnik:2011mv,Fok:2011yc}. We find the strongest constraint comes from the vector $\Upsilon_{+0}$, which decays dominantly to SM fermions including $\Upsilon_{+0}\to \ell \nu$, as shown in Fig.~\ref{f.QuirkDecays}. The pseudoscalar $\eta_{+0}$ instead decays mostly to $W\gamma$ and $WZ$, leading to weaker limits. 
The cross section for the resonant $\Upsilon_{\pm 0} \to \ell \nu$ signal is estimated as
\begin{equation} \label{eq:ellnuestimate}
2\,\sigma(pp \to  \overline{\psi}_+ \psi_{0} + \overline{\psi}_{0} \psi_-) \times r_{\Upsilon_{\pm 0}} \times \mathrm{BR} (\Upsilon_{\pm 0} \to \ell \nu)\,,
\end{equation}
where the overall factor $2$ accounts for the sum over the $B$ and $C$ sectors, $\mathrm{BR} (\Upsilon_{\pm 0} \to \ell \nu)$ is the branching ratio to one family of leptons, and $r_{\Upsilon_{\pm 0}}$ is the fraction of events that decay from the vector bound state at the end of the de-excitation process. A na\"ive estimate from simply counting the available degrees of freedom yields $r_{\Upsilon_{\pm 0}} = 3/4$. However, the production and de-excitation of quirks is unlikely to lead to a pure singlet or triplet state, but rather to a linear combination of the two. In this case the widths of both states affect the decay probability as
\begin{equation}
r_{\Upsilon_{\pm 0}} =  \frac{3 \Gamma(\Upsilon_{\pm 0})}{\Gamma(\eta_{\pm 0}) + 3 \Gamma(\Upsilon_{\pm 0})}\,,\label{e.chargedRatio}
\end{equation}
with $\Gamma(X)$ the total width of $X$. 
\begin{figure}[t]
\begin{center}
\includegraphics[width=0.65\textwidth]{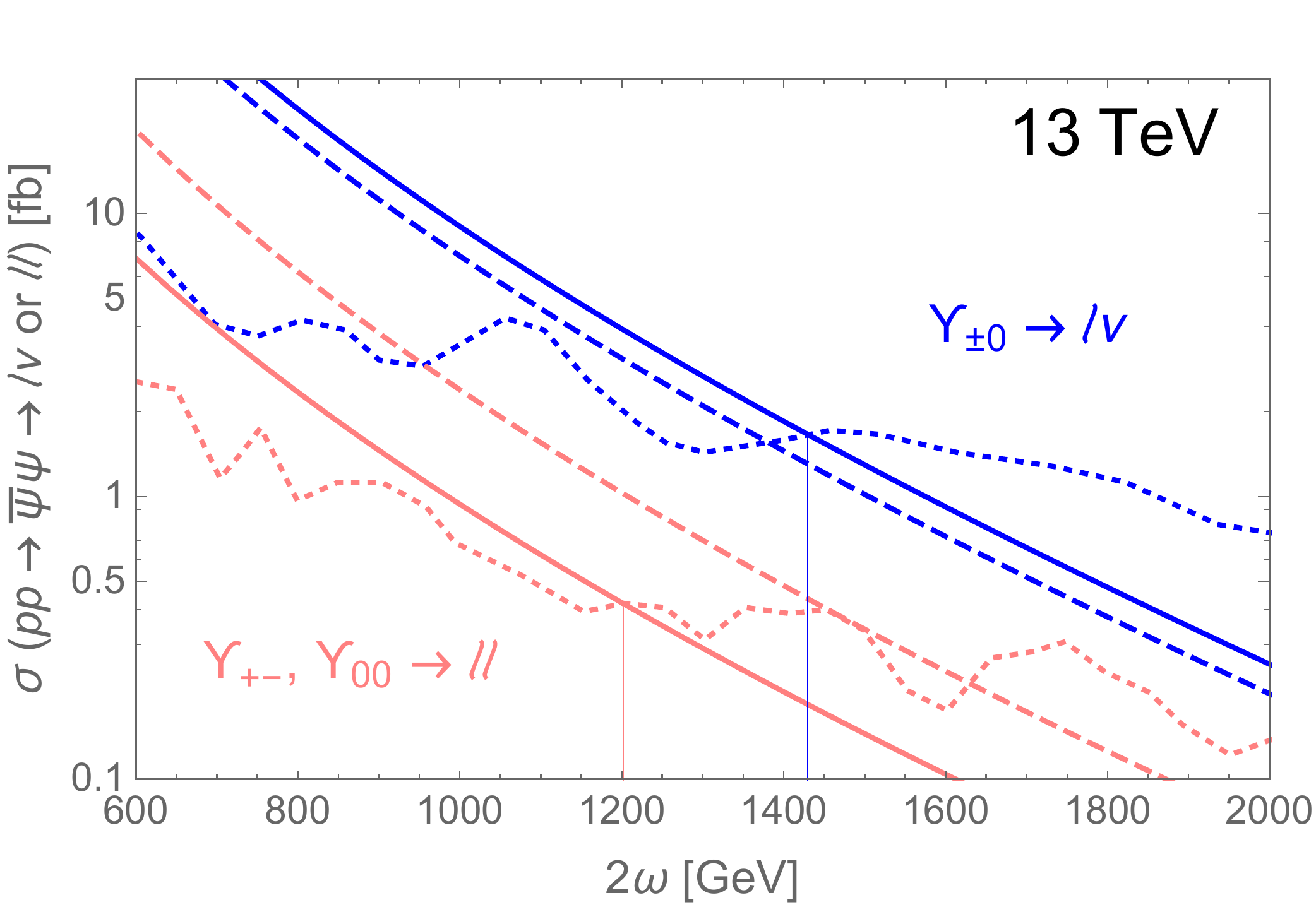}
\end{center}
\caption{Comparison of resonant quirkonium signals in the $\ell \nu$ (blue) and $\ell \ell$ (red) channels to the experimental bounds. Solid lines indicate the theory predictions, where the probability to decay from the vector bound state was computed according to Eq.~\eqref{e.chargedRatio} or its analogue for the electrically neutral bound states. Dashed lines show the effect of changing this probability to $3/4$. Dotted lines correspond to the current ATLAS $95\%$ CL cross section limits. The resulting lower bounds on the quirkonium mass are shown by the vertical lines.}
\label{f.quirkDecays}
\end{figure}
The multiplicity of SM fermion-antifermion final states makes $\Gamma(\Upsilon_{\pm 0})$ nearly $7$ times larger than $\Gamma(\eta_{\pm 0})$, yielding $r_{\Upsilon_{\pm 0}}\simeq 0.95$. Notice that this estimate is affected by a small nonperturbative uncertainty due to the unknown ratio of the wavefunctions at the origin of $\Upsilon$ and $\eta$, which we assume to be $1$. In Fig.~\ref{f.quirkDecays} we compare the signal cross section computed using Eq.~\eqref{eq:ellnuestimate} to the current  ATLAS bound~\cite{Aaboud:2017efa}. The two estimates for $r_{\Upsilon_{\pm 0}}$, namely $3/4$ and Eq.~\eqref{e.chargedRatio}, lead to similar limits $\omega \gtrsim 700\;\mathrm{GeV}$. Using the Coulomb approximation to evaluate the wavefunction at the origin, we find the total widths of $\Upsilon_{\pm 0}$ and $\eta_{\pm 0}$ are in the $1\,$-$\,10\;\mathrm{MeV}$ range. This corresponds to $t_{\rm ann} \lesssim 10^{-21}\;\mathrm{s} \ll t^{\,\gamma}_{\,\text{de-excite}}\,$, confirming that annihilation takes place immediately once the system reaches its ground state.

The electrically neutral quirkonia $\overline{\psi}_+ \psi_-$ and $\overline{\psi}_0 \psi_0$ are produced in DY via $Z$ and photon exchange. For $\omega = 500\;\mathrm{GeV}$ the $13\;\mathrm{TeV}$ production cross sections are $17\;\mathrm{fb}$ and $15\;\mathrm{fb}$, respectively. In contrast to the charged case, the neutral pseudoscalars $\eta_{+-}$ and $\eta_{00}$ decay dominantly to two hidden gluons, which in turn hadronize into glueballs. While this may lead to observable displaced decays in ATLAS and CMS~\cite{Chacko:2015fbc}, as discussed after Eq.~\eqref{e.glueballctau} the estimate of the glueball lifetime suffers from large theoretical uncertainties. A more robust signature is the dileptonic decay of $\Upsilon_{+-}$ and $\Upsilon_{00}$, whose rate is given by a formula similar to Eq.~\eqref{eq:ellnuestimate}. The large hadronic widths of the pseudoscalars imply a suppression of the probability to decay from the vector states: from the analogs of Eq.~\eqref{e.chargedRatio} we find $r_{\Upsilon_{+-}} \simeq r_{\Upsilon_{00}} \simeq 0.30$. In addition, numerically $\mathrm{BR}(\Upsilon_{+-} \to \ell \ell)/\mathrm{BR}(\Upsilon_{00} \to \ell \ell) \simeq 2.4$, hence the signal from the $\Upsilon_{+-}$ dominates. The comparison of the total signal cross section to the current  ATLAS bound~\cite{Aaboud:2017buh} is shown in Fig.~\ref{f.quirkDecays}. The resulting limit is $\omega \gtrsim 600\;\mathrm{GeV}$, weaker than the one coming from the charged channel. The $\eta_{+-} \to \gamma\gamma$ decay also leads to $\omega\gtrsim 600\;\mathrm{GeV}$, as determined by comparing the signal prediction (enhanced by $r_{\eta_{+-}} \simeq 0.70$) to the experimental limits on diphoton resonances \cite{Aaboud:2017yyg}. 

The quirkonium bounds discussed above and illustrated in Fig.~\ref{f.quirkDecays} are robust when the siblings are heavier than the cousins, i.e., for $\Delta > \omega$. In this case, the presence of light EWinos or hidden gluinos may open new decay channels to these superpartners and therefore modify the quirkonium branching ratios, but the constraints on $\omega$ are not strongly altered. 
\begin{figure}[t]
\begin{fmffile}{EWino}
\begin{fmfgraph*}(120,80)
\fmfpen{1.0}
\fmfstraight
\fmfset{arrow_ang}{15}
\fmfset{wiggly_slope}{70}
\fmfset{wiggly_len}{5mm}
\fmfleft{p1,i1,p2}\fmfright{o1,p3,o2}
\fmf{fermion,tension=1.2}{i1,v1}
\fmf{dashes,tension=0.6}{v1,o1}
\fmf{fermion,tension=0.6}{v1,o2}
\fmffreeze
\fmf{boson,tension=1.0}{v1,o2}
\fmfv{l=$\psi_{\pm,,0}$}{i1}
\fmfv{l=$\tilde{s}^c_\Delta$,l.a=0}{o1}
\fmfv{l=$\tilde{\chi}^{\pm,,0}$,l.a=0}{o2}
\fmfv{decor.shape=circle,decor.filled=full,decor.size=1.5thick}{v1} 
\end{fmfgraph*}
\end{fmffile}
\caption{\label{f.DecayToEWino}Decay of a quirk to a light sibling and a light EWino in the case $\omega \gtrsim \Delta + m_{\tilde{\chi}}\,$.}
\end{figure}
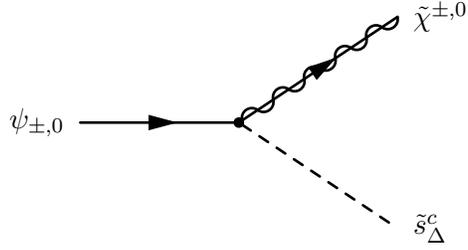

\begin{figure}[t]
 \begin{center}
\includegraphics[width=0.5\textwidth]{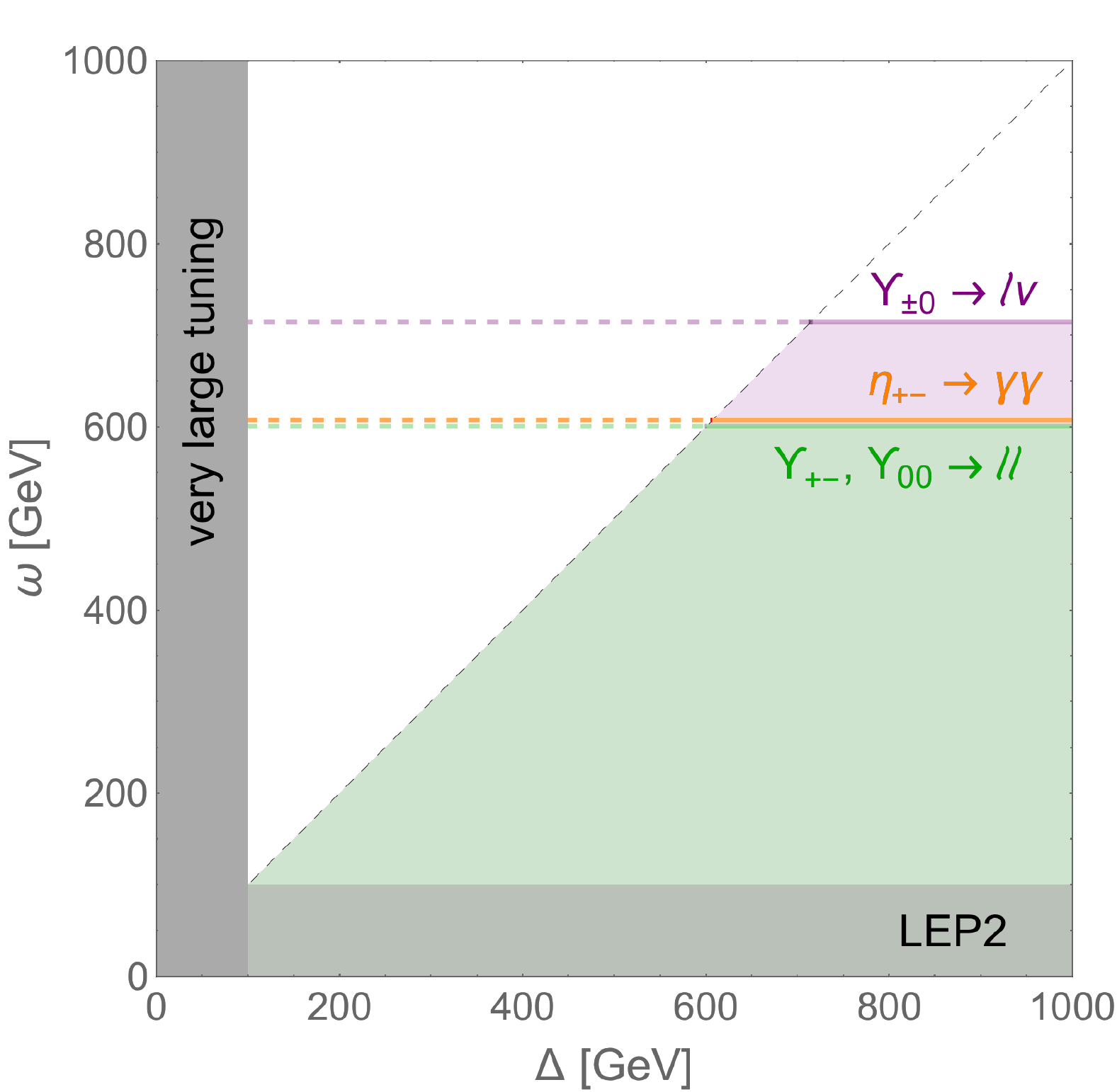} 
 \end{center}
 \caption{Summary of the current constraints on the parameter space. We do not consider the region $\Delta < 100\;\mathrm{GeV}$, where the fine tuning $\sim \Delta^2/M^2$ becomes very severe. Bounds from LEP2 rule out the gray shaded region $\omega< 100$ GeV. In purple, green, and orange we show the exclusions coming from the $\ell \nu, \ell \ell$ and $\gamma\gamma$ signals of the quirkonia, respectively, see Sec.~\ref{ssec.QuirkSig}. The quirkonium constraints can be relaxed or removed if $ \Delta < \omega $ and some EWinos are light.} 
\label{f.bosonDecays}
\end{figure}

In the opposite regime $\Delta < \omega$, if there are light EWinos the fermionic cousins can decay to a light sibling and an EWino, as depicted in Fig.~\ref{f.DecayToEWino}. The quirkonium annihilation signals are then erased and replaced by those of the light siblings, which behave as scalar quirks (``squirks''). Their phenomenology is discussed in Sec.~\ref{ssec.SquirkSigSibling}. A summary of the quirk constraints on the $(\Delta, \omega)$ parameter space is shown in Fig.~\ref{f.bosonDecays}.

\subsection{Cousin squirks\label{ssec.SquirkSigCousin}}
The phenomenology of the squirks shares several features with that of the fermions. The main production process is DY, and annihilation of squirky bound states is suppressed when $\ell > 0$. In contrast to the fermionic case, however, there is only one $s$-wave state $\chi$, with $0^{++}$ quantum numbers. Here we consider the scenario $\Delta > \omega$, when the siblings $\tilde{s}_\Delta^c$ are heavy and we can focus on the scalar cousins. The opposite regime $\Delta < \omega$ is discussed in Sec.~\ref{ssec.SquirkSigSibling}. 

Once a pair of scalar cousins is produced in quark-antiquark annihilation, it de-excites by radiating soft photons. Bound states of two electrically neutral squirks radiate through mixing with bound states composed of two charged squirks. The associated characteristic timescale is given by a formula analogous to Eq.~\eqref{e.mixingGamma}, but with $\Delta m_\psi$ replaced by the relevant splitting of the charged and neutral scalar masses. The largest splitting is the one among $\tilde{b}_B^{\prime c}$ and $\tilde{s}_\omega^c$, with the former heavier by $\approx \tfrac{1}{2} m_t^2 \omega / (\Delta^2 - \omega^2)$ which is typically of few tens of GeV, i.e., an order of magnitude larger than $\Delta m_\psi$. The resulting timescale for de-excitation via photon emission is $\sim 10^{-15}\;\mathrm{s}$, leading us to conclude that all scalar pairs promptly reach the ground state and annihilate.

The most promising signals arise from electrically charged squirkonia, that have larger DY cross section and cannot annihilate entirely to hidden glueballs. The production cross section is
\begin{equation}
\hat{\sigma}(u\bar{d} \to \tilde{s}\, \tilde{b}_B^\ast) \simeq \frac{\pi \alpha_W^2}{24\,\hat{s}} \frac{\hat{s}^2}{(\hat{s} - m_W^2)^2} \left(1 - \frac{4 \omega^2}{\hat{s}} \right)^{3/2},
\end{equation}
where we neglect the effects of the mixing between $\tilde{s}$ and $\widetilde{S}$. Numerically, for $\omega = 500\;\mathrm{GeV}$ we find $2\,\sigma(pp \to  \tilde{s}\, \tilde{b}_B^\ast  +  \tilde{s}^\ast \tilde{b}_B ) = 8.9\;\mathrm{fb}$ at $13\;\mathrm{TeV}$, where the factor $2$ includes also the production of $\tilde{b}_B^{\prime c} \tilde{s}^{c\, \ast}_\omega + \tilde{b}_B^{\prime c\,\ast } \tilde{s}^{c}_\omega $, again neglecting mixing effects. We find that the total production cross section of the cousin scalars is suppressed by a factor $\approx 6.7$ compared to their fermionic counterparts. The annihilation patterns differ for the two doublets. Before mixing with the singlet scalars, $\widetilde{Q}_B$ couples to the Higgs with $y_t$ strength, hence $ \tilde{s}\, \tilde{b}_B^\ast$ pairs annihilate dominantly to $Wh$, whereas $\widetilde{Q}_B^{\prime c}$ does not couple to the Higgs, so $\tilde{b}_B^{\prime c} \tilde{s}^{c\, \ast}_\omega$ pairs annihilate mostly to the $W\gamma$ and $WZ$ final states \cite{Burdman:2008ek,Burdman:2014zta}, with $\mathrm{BR}(W\gamma)/\mathrm{BR}(WZ) \simeq (\tan^{2}\theta_w)^{-1} \approx 3.3$. Mixings with the singlets give $O(1)$ modifications of the branching ratios, but are not expected to change this picture qualitatively. Thus, the most important constraints come from searches for $Wh$~\cite{Aaboud:2017cxo} and $W\gamma$~\cite{Aad:2014fha} resonances. From Ref.~\cite{Aaboud:2017cxo} we find the $Wh\to \ell \nu b\bar{b}$ cross section expected from the cousin squirks is just below the experimental bound in the $300\;\mathrm{GeV} \lesssim \omega \lesssim 500\;\mathrm{GeV}$ range, but currently no exclusion applies. On the other hand, the $W\gamma \to \ell \nu \gamma$ final state \cite{Aad:2014fha} yields a weak bound $\omega \gtrsim 300\;\mathrm{GeV}$, though this search has not been updated yet to $13\;\mathrm{TeV}$ data. 

Because the mass splitting between $\tilde{b}_B^{\prime c}$ and $\tilde{s}_\omega^c$ is $O(10) \times \Delta m_\psi$, Eq.~\eqref{eq:betadecayfermions} implies the timescale for beta decay is roughly $\sim 10^{-19}\;\mathrm{s}$, which is of the same order as de-excitation via photon emission, Eq.~\eqref{e.photonTime}. Thus, beta decay may occur before de-excitation, which would erase the $W\gamma/WZ$ resonant signals. On the other hand, $\tilde{b}_B$ and $\tilde{s}$ have the same mass splitting $\Delta m_\psi$ as the quirks, so they de-excite and annihilate to $Wh$ well before beta decay becomes effective. To summarize, for $\Delta > \omega$ the squirk phenomenology is subleading to that of the quirks.

\subsection{Light sibling squirks\label{ssec.SquirkSigSibling}}
The sibling scalar is dominantly a SM singlet. Its direct DY pair production only proceeds through mixing with the cousin scalars, and is therefore suppressed. However, if $\Delta < \omega$, as we assume in this subsection, the light siblings can also be produced by the decays of cousin fermions and scalars. The fermions $\psi_{\pm, 0}$ can decay to $\tilde{s}_\Delta^c$ and a light EWino, if kinematically allowed, see Fig.~\ref{f.DecayToEWino}. The scalar cousins, on the other hand, decay to $\tilde{s}_\Delta^c$ and a gauge boson or Higgs via mass mixing. For $\tilde{s}$ and $\tilde{b}_B$, the mixing is mediated by the subleading A- and B-terms in Eq.~\eqref{eq:extraAandBterms}. Thus, cousin pair production typically results in a $\tilde{s}_\Delta^{c \,\ast} \tilde{s}_\Delta^c$ squirky bound state. Due to the singlet nature of the siblings, the de-excitation of this system is a complex process, which we now analyze in detail.

Photon radiation via mixing with the $\tilde{b}_B^{\prime c \ast} \tilde{b}_B^{\prime c}$ bound state is strongly suppressed by the large $\omega - m_{\tilde{s}_\Delta^c}$ mass splitting. For $\Delta \ll \omega$ this de-excitation timescale is approximately given by Eq.~\eqref{e.mixingGamma} with $\Delta m_\psi$ replaced by $\sim \omega$, and multiplied by $(\sin \phi_R)^{-4}$ due to the reduced coupling of $\tilde{s}_\Delta^c$ to the $W$ boson. The resulting lifetime of about $10^{-10} \;\mathrm{s}$ shows that photon radiation is very slow.  

Turning to glueball radiation, we adapt the arguments of Ref.~\cite{Kang:2008ea} to scalar constituents. This implies that the kinematic suppression may be as large as $(\sqrt{\sigma}/m_0)^8$, leading to a timescale $\lesssim 10^{-16}\;\mathrm{s}$, which is still prompt. Just like quirkonium, glueball radiation typically does not reach the ground state, leaving a residual kinetic energy $K\lesssim m_0$. The de-excitation can be completed by radiating light SM fermions via an off-shell $Z$ boson. The corresponding timescale is estimated by applying the photon radiation formula, Eq.~\eqref{e.photonTimeGen}, with $K \sim m_0$ and the replacement
\begin{equation}
\alpha\;\to \; \frac{\alpha_W^2\sin^4\phi_R N_f}{ 4\pi\, 4\cos^4 \theta_w}\left(\frac{\delta E}{m_Z}\right)^4 .
\end{equation}
This effective interaction strength takes into account the $\sin^2\phi_R$ suppression of the $Z$-$\tilde{s}_\Delta^{c \ast}$-$\tilde{s}_\Delta^{c}$ coupling and includes the multiplicity $N_f$ of kinematically available SM fermions. The powers of $m_Z$ originating from the $Z$ propagator are compensated by the typical splitting $\delta E$ between adjacent energy levels. Averaging over the differences between the $E_n$ in Eq.~\eqref{e.energystates} (with $\mu = m_{\tilde{s}^c_\Delta}/2$) below the glueball mass, we obtain
\begin{equation} \label{eq:deltaE}
\delta E\approx \frac{3\pi}{2}\frac{\sigma}{\sqrt{m_0 m_{\tilde{s}_\Delta^c}}}\,.
\end{equation}
Combining the different pieces we arrive at the result
\begin{equation}
t^{\,Z}_{\,\text{de-excite}}\sim\frac{32}{27 \pi^4} \frac{\cos^4 \theta_w }{\alpha_W^2 \sin^4 \phi_R N_f} \frac{ m_Z^4 m^4_{\tilde{s}_\Delta^c} m_0^3}{\sigma^6} \sim 4 \cdot  10^{-13}\;\text{s}\,\left(\frac{5 \,\text{GeV}}{\Lambda_{\text{QCD}_{B,C}}} \right)^{9}\left(\frac{m_{\tilde{s}_\Delta^c}}{300\,\text{GeV}} \right)^{4}, 
\end{equation}
where in the numerical estimate we take $\sin \phi_R = 0.4$ and $N_f = 18$ as typical values. This corresponds to a proper decay length of $\sim 0.1\;\mathrm{mm}$, for which the sensitivity of LHC displaced decay searches is severely degraded. Furthermore, our lifetime estimate is conservative. In using the Larmor formula in Eq.~(\ref{e.larmor}) we modeled the de-excitation as a sequence of small transitions of energy $\delta E \ll K \sim m_0$. This is a semi-classical picture, and in fact Eq.~\eqref{eq:deltaE} may be regarded as the classical radiation frequency. However, we should also consider direct transitions to the ground state, whose amplitudes are suppressed by wavefunction overlap integrals, but enhanced by the larger $\delta E \sim m_0$. For example, a dipole transition between an excited state with $K\sim m_0$ and $\ell = 1$ and the ground state, has a lifetime of $\sim 10^{-16}\;\mathrm{s}$, several orders of magnitude shorter than the multi-step process described above. Higher multipoles with $\Delta \ell \sim \mathrm{few}$ can also accelerate the de-excitation, despite being more suppressed. In conclusion, we expect that the $\tilde{s}_\Delta^{c\ast} \tilde{s}_\Delta^c$ system de-excites promptly to its ground state and annihilates, and proceed under this assumption. 

\begin{figure}[t]
\includegraphics[width=0.45\textwidth]{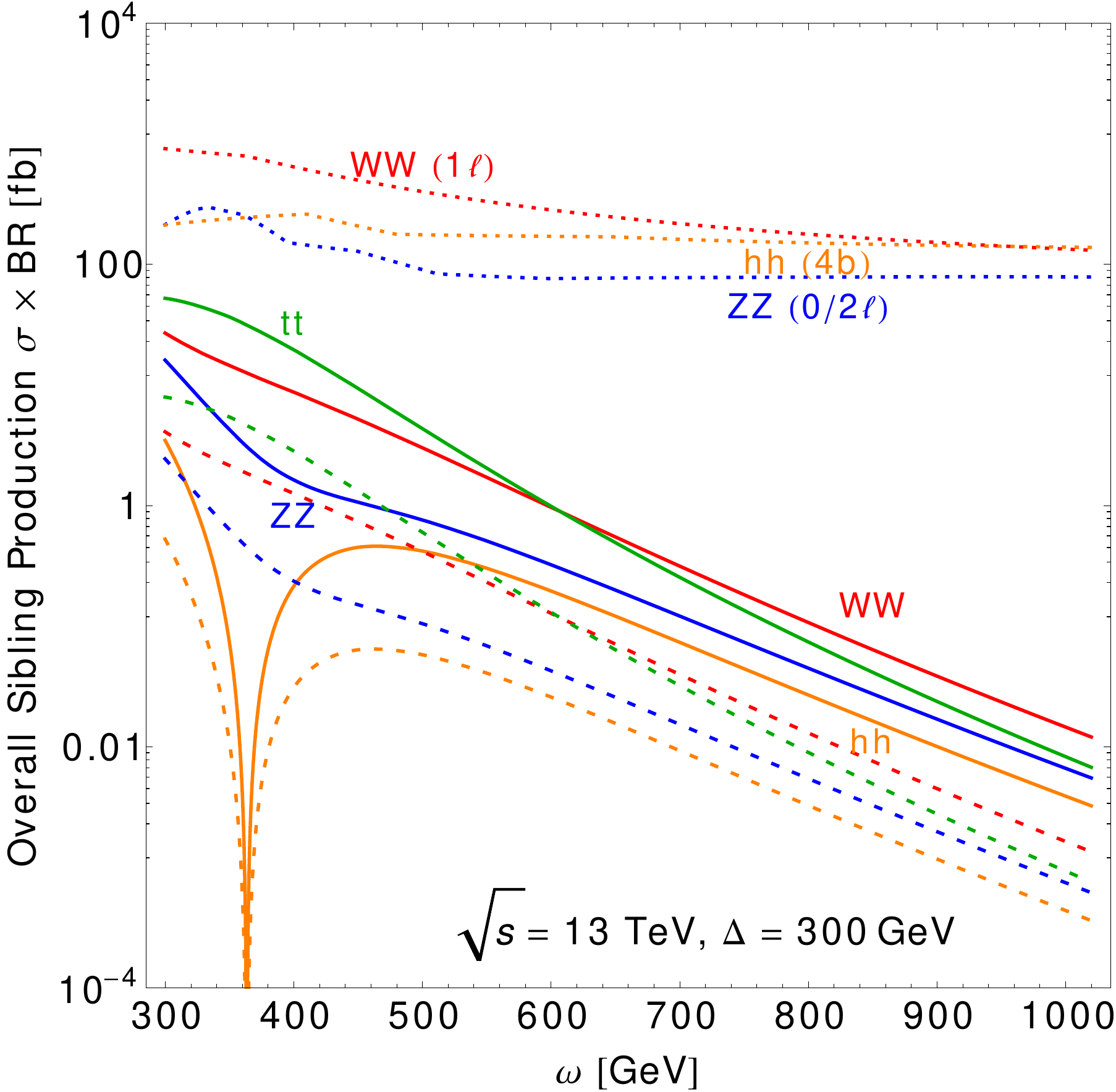}\hspace{4mm}
\includegraphics[width=0.45\textwidth]{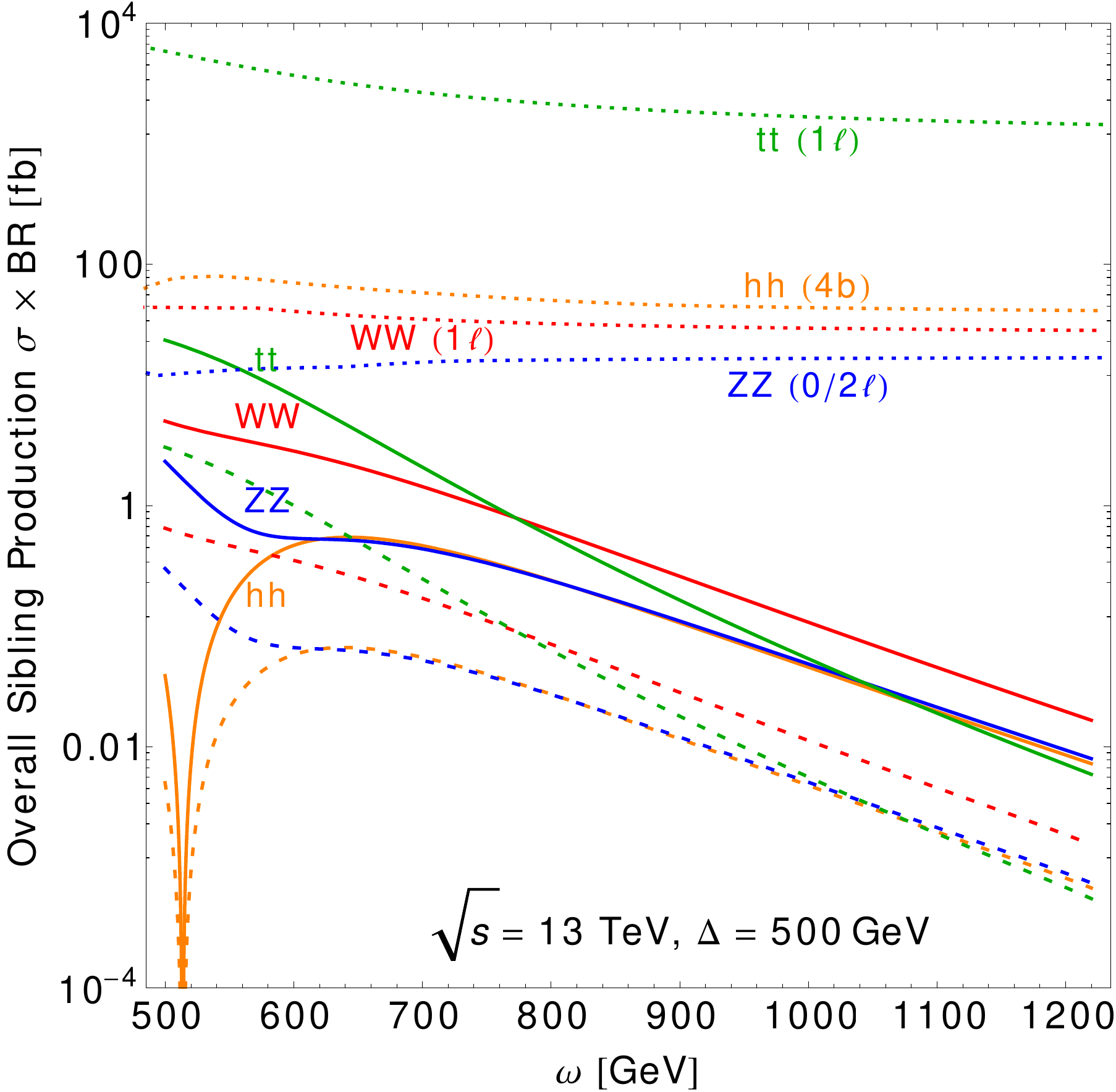}
\caption{Cross sections for sibling squirkonium production at $13$ TeV times branching ratios for decay to SM particles, vs. the experimental bounds. We set $\Delta=300\,(500)$ GeV in the left (right) panel. Red, blue, orange, and green curves correspond to the $WW,ZZ, hh$, and $t\bar{t}$ channels, respectively.  Solid curves assume that all cousins, both fermions and scalars, decay to the light sibling, whereas dashed curves include only the contributions from all squirk pairs. Dotted curves show the current experimental limits on resonances with mass $2 m_{\tilde{s}_\Delta^c} (\lesssim 2\Delta)$. The $t\bar{t}$ constraint is too weak to appear in the left panel.}
\label{fig:NeuBoundMax_Mix}
\end{figure}

Since $\tilde{s}_\Delta^c$ is mostly a SM singlet, the lowest-lying bound state $\chi_{\Delta \Delta}$ annihilates dominantly to hidden glueballs. As mentioned earlier, signals from glueball displaced decays~\cite{Chacko:2015fbc} are possible but not guaranteed, given the large uncertainty in the lifetime prediction of Eq.~\eqref{e.glueballctau}. A robust signature is provided by the subleading decays to SM dibosons, $\chi_{\Delta \Delta}\to WW, ZZ, hh$, and fermions, $\chi_{\Delta\Delta}\to t\bar{t}$, which arise due to the mass mixing in Eq.~\eqref{eq:M2Sc}. The branching fractions of these modes are, however, at most a few percent due to the large hidden QCD coupling, $\alpha_d (m_{\tilde{s}_\Delta^c}) \sim 0.1$ for sibling masses of few hundred GeV and $\Lambda_{\mathrm{QCD}_{B,C}} = 5\;\mathrm{GeV}$. Assuming that all the cousins, both fermions and scalars, decay to the light sibling plus additional particles, the total production cross section of $\tilde{s}_\Delta^{c\ast}\tilde{s}_\Delta^{c}+\mathrm{anything}$ is obtained by summing the production cross sections of all siblings and cousins. The dominant contribution comes from the production of the cousin fermions $\psi_\pm, \psi_0$. In Fig.~\ref{fig:NeuBoundMax_Mix} we show the total cross section multiplied by the branching ratios of $\chi_{\Delta \Delta}$ decays in various SM channels, which are given in Appendix~\ref{app:squirkonium}. The comparison with the experimental bounds on resonances with mass $2 m_{\tilde{s}_\Delta^c} (\lesssim 2\Delta)$ in the $WW$~\cite{Aaboud:2017fgj}, $ZZ$~\cite{Aaboud:2017itg}, $hh$~\cite{Aaboud:2016xco}, and $t\bar{t}$~\cite{ATLAS-CONF-2016-014} final states shows that the $\chi_{\Delta\Delta}$ signals are at least an order of magnitude below the current sensitivity.

We note that additional particles produced along with the sibling pair can potentially lead to further constraints. For example, pair production of the charged quirks $\overline{\psi}_+ \psi_-$ can yield the final state $\chi_{\Delta \Delta} + W^+W^- + \tilde{\chi}^0 \tilde{\chi}^0$, resulting in $W^+ W^- +$ missing transverse energy (MET) if the sibling pair annihilates to invisible glueballs, or a multi-gauge-boson + MET final state if $\chi_{\Delta\Delta}\to WW$ or $ZZ$. Many other possibilities exist, but the dominant signals and the associated bounds depend strongly on the spectra of the EWinos and the hidden sector. We therefore defer the study of these signatures to future work. 

Finally, we comment on the $\Delta \sim \omega$ region. Even if the lightest neutralino is very light, when the mass splitting between the cousin fermions and the sibling is small, $\omega - m_{\tilde{s}_\Delta^c} \ll m_W$, the $4$-body decay of the charged quirk $\psi_- \to \tilde{s}_\Delta^c \tilde{\chi}^{-\ast}\to \tilde{s}_\Delta^c \tilde{\chi}^0 (W^{- \ast} \to f \bar{f}^\prime)$ can be slower than the de-excitation via photon emission, whose timescale was given in Eq.~\eqref{e.photonTime}. In this case, the quirk pair can annihilate to SM particles as discussed in Sec.~\ref{ssec.QuirkSig}. When kinematically allowed, a quirk pair containing one or two neutral quirks may convert into a squirk pair by exchanging a hidden gluino, although this process is unlikely to dominate due to the mixing angle suppression. Nevertheless, annihilation signals that originate from $\overline{\psi}_+ \psi_-$ pairs should survive and lead to significant bounds on $\omega$ from the $\Upsilon_{+-}\to \ell\ell$ and $\eta_{+-} \to \gamma\gamma$ decay channels. Similarly, for small mass splitting the decays of the scalar cousins to $\tilde{s}_\Delta^c$ and (off-shell) gauge or Higgs bosons can become very suppressed, and thus ineffective in preventing the annihilation of squirky cousin pairs. However, for $\Delta$ even moderately smaller than $\omega$ the decays dominate, and all cousins cascade down to the light sibling.

\section{Conclusions\label{sec:conclusion}}
The lack of LHC signals from new colored particles motivates the broad framework of neutral naturalness. In this article we presented the first supersymmetric model where the top partners are complete SM singlet scalars, which we dubbed top siblings. While inspired by Folded SUSY, our construction differs from it in several aspects. It is purely four dimensional, thus allowing enough parametric freedom to easily accommodate realistic electroweak symmetry breaking. Two hidden top sectors are needed to cancel the quadratic top-loop corrections to the Higgs mass, but no hidden light generations are necessary. The model also requires that the soft masses of the colored stops and of the siblings are equal in magnitude, but opposite in sign. We have provided an explicit construction that realizes this structure, where the top superfields, both visible and hidden, arise as IR composite degrees of freedom of strongly coupled SUSY gauge theories. The associated UV cutoff can be as high as $100\;\mathrm{TeV}$, an order of magnitude larger than in many neutral naturalness models. 

Probing directly the SM-singlet siblings is a challenge for the LHC experiments. Consequently, the collider phenomenology is largely governed by the top cousins, which are electroweak-charged fermions and scalars that accompany the siblings. When the sibling mass $\Delta$ is larger than the cousin mass $\omega$, the resonant annihilation signals of the cousin quirks lead to a bound $\omega \gtrsim 700\;\mathrm{GeV}$. However, in the opposite regime $\Delta < \omega$ these constraints can be relaxed, or altogether erased, if the cousins rapidly decay down to the light siblings. This happens if at least some of the EWinos are light. Then, the electroweak pair production of the cousins results in the formation of squirky pairs of siblings. These annihilate dominantly to hidden glueballs, which are relatively long lived in our model and can decay outside the LHC detectors. Annihilation to SM particles is very suppressed, so very light siblings are currently compatible with LHC constraints.

Several interesting variations of our model can be envisaged. One possibility is to give the siblings nonzero hypercharge, which results in different patterns of electroweak signals. Another appealing option is to switch the roles of $SU(2)_L$ singlets and doublets of the hidden sectors, $ u_{B,C}^c  \leftrightarrow Q_{B,C} $ and $ u^\prime_{B,C}  \leftrightarrow Q_{B,C}^{\prime c} $, both in the superpotential and in the soft masses. The Higgs potential is unaffected by this transformation. In this alternative version of the model the top partners are EW doublet scalars, whereas the cousins are complete SM singlets. Since it is technically natural for their mass $\omega$ to be small, the cousins could be very light, leading to exotic phenomenology. These possibilities will be investigated in a future publication.

%%%%%%%%%%%%%%%%%%%%%%%%%%%%%%%%%%%%%%%
\acknowledgments
We thank Nathaniel Craig, Simon Knapen, Markus Luty, Riccardo Rattazzi, and John Terning for helpful discussions. H.-C.C., L.L., and C.B.V. are supported by Department of Energy Grant number DE-SC-000999. H.-C.C. was also supported by The Ambrose Monell Foundation at the Institute for Advanced Study, Princeton. C.B.V. thanks the TUM Physics Department for hospitality during the completion of this work. E.S. has been partially supported by the DFG Cluster of Excellence 153 ``Origin and Structure of the Universe,'' the Collaborative Research Center Grant No.~SFB1258 and the COST Action Grant No.~CA15108, and is grateful to the CERN Theory group for hospitality and partial support in the final stages of this project.

\appendix
\section{Soft masses of composite mesons}
\label{s.softmasses}
The relations between the soft SUSY-breaking masses of UV constituents and IR composites were derived in Ref.~\cite{ArkaniHamed:1998wc}. In this appendix we briefly summarize those relations and generalize the result to the case of non-universal soft masses. Consider a SUSY gauge theory with a ``quark'' $P$ transforming under the gauge group in the UV. Under the reparametrization of superfield $ P \to \sqrt{Z} P$, the rescaling anomaly generates a shift in the holomorphic gauge coupling function $S= \frac{1}{ g^2} - i \frac{\theta}{8\pi^2}$:
\begin{equation}
S(\mu_\text{UV}) \to S(\mu_\text{UV}) + \frac{T}{8\pi^2} \ln Z,
\end{equation}
where $T$ is the Dynkin index of the representation under which $P$ transforms. The Lagrangian can be written as
\begin{equation}
\frac{1}{4} \int d^2\theta\, S(\mu_\text{UV}) W^2 + \text{h.c.} + \int d^4\theta  Z F\left( S(\mu_\text{UV})+ S^\dagger(\mu_\text{UV})- \frac{T}{4\pi^2} \ln Z\right) P^\dagger e^VP,
\end{equation}
where the first term is the gauge kinetic term. The theory is invariant under the transformation
\begin{equation}
Z \to Z \chi \chi^\dagger, \quad P \to P/\chi, \quad S(\mu_\text{UV}) \to S(\mu_\text{UV}) + \frac{T}{4\pi^2} \ln \chi \,.
\label{e.u1A}
\end{equation}

The functions $S(\mu_\text{UV})$ and $Z$ can be promoted to chiral and real vector superfields respectively. One can see that $\ln Z$ plays the role of the background vector superfield of the anomalous $U(1)_A$ symmetry of Eq.~(\ref{e.u1A}). Physical quantities must be invariant under $U(1)_A$ and renormalization group transformations. The only such quantity that can be formed from $S$ and $Z$ is
\begin{equation}
I  = \Lambda_h^\dagger Z^{2T/b} \Lambda_h ,
\label{e.strong_scale}
\end{equation}
where $\Lambda_h = \mu_\text{UV} e^{-8\pi^2 S/b}$ is the holomorphic dynamic scale and $b$ is the beta function coefficient. If there is more than one quark transforming under the gauge group, the expression generalizes to
\begin{equation}
I  = \Lambda_h^\dagger \left(\prod_k Z_k^{2T_{r_k}/b}\right) \Lambda_h ,
\label{e.strong_scale2}
\end{equation}
where $T_{r_k}$ is the Dynkin index of the representation $r_k$ for the $k$th quark. The physical strong scale corresponds to the $\theta^0$ component of $I$, $\Lambda^2 = [I]_{\theta=\bar{\theta}=0}$. We may view Eq.~(\ref{e.strong_scale2}) as indicating that $I$ carries charges $2T_{r_k}/b$ under  $U(1)_{A,k}$.

The $\theta^2$ component of $S$ and $\theta^2 \bar{\theta}^2$ component of $\ln Z$ correspond to the soft gaugino and squark masses respectively. In particular,
\begin{equation}
m_P^2 (\mu_\text{UV}) = - [\ln Z]_{\theta^2 \bar{\theta}^2} - [\ln F(\mu_\text{UV})]_{\theta^2 \bar{\theta}^2} \stackrel{\mu_\text{UV} \to \infty}{\xrightarrow{\hspace*{1.2cm}}}  - [\ln Z]_{\theta^2 \bar{\theta}^2}\,,
\end{equation}
as the contribution from $F$ is proportional to the anomalous dimension and its derivative~\cite{ArkaniHamed:1998wc} which vanish in the $\mu_\text{UV} \to \infty$ limit for an asymptotically free theory.

In an s-confining theory, the low-energy degrees of freedom are mesons $M$ and baryons $B, \overline{B}$. Near the origin, the K\"ahler potential can be expanded in power series in $M, B,$ and $\overline{B}$. For the meson field $M_{ij}$ made of $P_i$ and $\overline{P}_j$, its K\"ahler potential must start with
\begin{equation}
K \supset  c_{M_{ij}} \frac{ M^\dagger_{ij} Z_i Z_{\bar{j}} M_{ij}}{I} + \cdots
\end{equation}
to have the correct dimension and be invariant under the $U(1)_A$ symmetries. The leading soft SUSY-breaking mass of the meson $M_{ij}$ can be similarly obtained. In the far IR where the contribution from $c_{M_{ij}}$ vanishes (similar to that of the $F$ function in the UV), we have 
\begin{align}
m^2_{ M_{ij}} = - \left[ \ln \frac{Z_i Z_{\bar{j}}}{I} \right]_{\theta^2 \bar{\theta}^2}  =& -[\ln Z_i ]_{\theta^2 \bar{\theta}^2} - [\ln Z_{\bar{j}}]_{\theta^2 \bar{\theta}^2} + [\ln I]_{\theta^2 \bar{\theta}^2}\\
 =& \: m^2_{P_i}+m^2_{\overline{P}_j}- \frac{2 }{b}\sum_k T_{r_k}\left(m^2_{P_k}+m^2_{\overline{P}_k}\right), 
\end{align}
which is Eq.~\eqref{e.mesonMass}. 

\section{Estimating $\Lambda_{\text{QCD}_{B,C}}$\label{a.lambdaQCD}}
In this appendix we describe our estimation of $\Lambda_{\text{QCD}_{B,C}}$. At the high scale $\Lambda_{Z_3}$ the strong coupling constants of the $A$, $B$, and $C$ sectors are assumed to be equal, but the different particle content of the visible and hidden sectors result in different coupling values at lower energies, $\alpha_A (\mu) \neq \alpha_{B,C} (\mu)$, and therefore different confinement scales. Since the phenomenology is very sensitive to the value of $\Lambda_{\text{QCD}_{B,C}}$, we perform the RG running at two loops. At this order we have~\cite{Caswell:1974gg,Jones:1974pg,Machacek:1983tz}
\begin{equation}
\frac{d \alpha_i}{d \ln \mu^2}=-\,\frac{\alpha_i^2}{4\pi} \left( b + b_1 \frac{\alpha_i}{4\pi} + O(\alpha_i^2) \right),
\end{equation}
for $i = A, B, C$, with
\begin{align}
b\,=\,& 11 - \frac{1}{3} \,n_f - \frac{1}{6} \,n_s - 2 n_{\tilde{g}}\, ,\\
b_1\,=\,& 102 - \frac{19}{3}\, n_f - \frac{11}{3} \,n_s - 48 n_{\tilde{g}} + \frac{13}{3}\, n_{\tilde{g}} \,\text{min}\left(n_f,n_s \right)  .
\end{align}
In these formulae $n_f$ and $n_s$ are the number of Weyl fermions and complex scalars, respectively, transforming as fundamentals of $SU(3)_i$. The number $n_{\tilde{g}}$ is either $1$ or $0$, depending on whether or not the gaugino is active in the running. Finally, the last term in $b_1$ arises from the SUSY gluino-fermion-scalar interactions. These interactions contribute only for complete active SUSY multiplets, whose number we count using the minimum function. The running between two thresholds $\mu_1$ and $\mu_2$ is determined by
\begin{equation}
\alpha^{-1}_i(\mu_1)-\alpha^{-1}_i(\mu_2) + B_1\ln \left(\frac{\alpha^{-1}_i (\mu_2) + B_1}{\alpha^{-1}_i(\mu_1) + B_1} \right) = B \ln\frac{\mu^2_1}{\mu^2_2}\,,
\end{equation}
where $B = b / (4\pi)$ and $B_1 = b_1 / (4\pi b)$. 
The confinement scale is (see e.g. Ref.~\cite{Ellis:1991qj})
\begin{equation}\label{e.confscale}
\frac{\Lambda_{\text{QCD}_{i}}}{\mu}=\exp\left( \frac{-1}{2 B \alpha_i(\mu)}\right) \left(\frac{1}{B \alpha_i(\mu)}+\frac{B_1}{B} \right)^{\frac{B_1}{2 B}} .
\end{equation}
Given a set of input parameters $M, \omega, \Delta$ and $m_{\tilde{g}_{B,C}}$, to determine $\Lambda_{\text{QCD}_{B,C}}$ we proceed as follows. Starting from $\alpha_A (m_Z) = 0.1185$ with $m_Z = 91.1876\;\mathrm{GeV}$, we run the $A$ coupling up to $\Lambda_{Z_3}$, taking into account the thresholds given by the top quark with $m_t = 173\;\mathrm{GeV}$, the $\tilde{t}_A, \tilde{b}_A$, $\tilde{u}_A^c$ with mass $\sim \sqrt{M^2 - \Delta^2}$, and the $A$ gluino and light generation squarks, whose masses are taken to be $M$. Then the $B,C$ couplings are run down according to the mass spectrum of the hidden sectors. Below the mass of the lightest hidden particle, Eq.~\eqref{e.confscale} is used to determine $\Lambda_{\text{QCD}_{B,C}}$.

In a more complete theory that explains the origin of the opposite sign soft masses, additional thresholds can be present. In the example discussed in Sec.~\ref{s.construct}, these are given by the confinement scale $\Lambda_G$ and by the masses of the extra composite mesons in the visible sector, $M_X$, and in the hidden sectors, $M_X^{B,C}$. As long as $M_X \sim M_X^{B,C}$, the additional thresholds have a mild effect on $\Lambda_{\text{QCD}_{B,C}}$.

\section{Mixing of electrically neutral quirkonia\label{a.QuirkoniumMixing}}
We describe the mixing between the two electrically neutral mesons by writing an effective Hamiltonian matrix in the $(Q_{00}, Q_{+-})^T$ space (see for example Ref.~\cite{Burdman:2003rs} for a review),
\begin{equation}
\bm{\mathrm{H}}=\bm{\mathrm{M}}_{0} + \bm{\mathrm{M}}_{\rm mixing} - \frac{i}{2}\, \bm{\Gamma} = 2\omega \left(\begin{array}{cc}
1-\frac{\Delta m_\psi}{\omega}&0\\
0&1
\end{array} \right) + \left(\begin{array}{cc}
0 & M_{\rm mix}\\
M_{\rm mix} &0
\end{array} \right) - \frac{i}{2} \left(\begin{array}{cc}
0&0\\
0&\Gamma_\gamma
\end{array} \right).
\end{equation}
Here $\bm{\mathrm{M}}_0$ contains the diagonal masses of the constituent quirks, which are split by $\Delta m_\psi$ as defined in Eq.~\eqref{e.quirksplitting}. Next, $\bm{\Gamma}$ contains the width of $Q_{+-}$ for photon emission, obtained from Eq.~\eqref{e.photonTime} as $\Gamma_\gamma = 1/t^{\,\gamma}_{\,\text{de-excite}}$. For our representative choice of the parameters, we have $\Gamma_\gamma \sim 3\;\mathrm{keV}$. Finally, $\bm{\mathrm{M}}_{\rm mixing}$ describes the mixing of the two mesons, 
\begin{equation}
M_{\rm mix} = \frac{1}{4\omega}  \langle Q_{00}|\mathcal{H}_\text{W} | Q_{+-} \rangle
\end{equation}
where $\mathcal{H}_\text{W} $ is the interaction Hamiltonian generated by $t$-channel $W$ exchange. The amplitude for this process is
\begin{equation}
- \frac{g^2}{2 (t - m_W^2)}\, \overline{u}_- \gamma^\mu u_0 \, \overline{v}_0 \gamma_\mu v_+ \,,
\end{equation}
where the contribution of the longitudinal $W$ was neglected, because it is relatively suppressed by $(\Delta m_\psi)^2/m_W^2$. To understand the expected size of $t$, recall that we are considering a transition from a highly excited state of $Q_{00}$ to one of $Q_{+-}$. The typical energy splitting among these highly excited states is of $O(\Lambda_{\text{QCD}_{B,C}})$, so we take $\sqrt{ -t } \sim \Lambda_{\text{QCD}_{B,C}} \ll m_W$. Then, by performing a Fierz rearrangement and neglecting an $O(1)$ coefficient, we obtain
\begin{equation}
\mathcal{H}_\text{W}\sim \frac{g^2} {2m_W^2}\,\mathcal{O}\,,
\end{equation}
where $\mathcal{O}$ is a $4$-fermion operator whose Lorentz structure depends on the type of bound states under consideration (e.g. scalar, vector, etc.). We then obtain 
\begin{equation}
M_{\rm mix} \sim \frac{1}{4\omega} \frac{g^2}{2 m_W^2}\langle Q_{00} |\mathcal{O} | Q_{+-} \rangle \sim \frac{1}{4\omega} \frac{g^2}{2 m_W^2}\, (2\omega)^2\Lambda_{\text{QCD}_{B,C}}^2 = 2\omega\, \frac{\Lambda_{\text{QCD}_{B,C}}^2}{v^2}\,,
\end{equation}
where again $O(1)$ factors were neglected. The decay width into photons of the mostly-$Q_{00}$ eigenstate can be extracted from the corresponding eigenvalue $\Omega_{00}$ as $\Gamma_{\gamma}^{00} = - 2 \,\mathrm{Im}\, \Omega_{00}$. Expanding to leading order in $\Gamma_\gamma$ we arrive at
\begin{equation}
\Gamma_{00}^\gamma \sim \frac{\omega^2 \Lambda_{\text{QCD}_{B,C}}^4}{(\Delta m_\psi)^2 v^4}\, \Gamma_\gamma\,,
\end{equation}
which can be rewritten as in Eq.~\eqref{e.mixingGamma}. Notice that since $\Gamma_\gamma$ is much smaller than all the other scales in the problem, this result can be simply obtained as $\Gamma_{00}^\gamma \sim \Theta^2 \Gamma_\gamma$, where $\Theta \sim M_{\rm mix} / (2 \Delta m_\psi) \sim \omega \Lambda_{\text{QCD}_{B,C}}^2 / (\Delta m_\psi \,v^2)$ is the mixing angle computed by neglecting $\Gamma_\gamma$ in the Hamiltonian.

\section{Annihilation of the light sibling squirkonium}\label{app:squirkonium}
The $\tilde{s}_\Delta^{c\ast} \tilde{s}_\Delta^c$ bound state $\chi_{\Delta\Delta}$ annihilates dominantly to $g_{B} g_{B}$, with subleading modes given by $t\bar{t}, WW, ZZ$ and $hh$. The corresponding decay widths can be obtained by adapting the results of e.g. Ref.~\cite{Martin:2008sv}. We have\footnote{In this appendix we take $\omega > \sqrt{\Delta^2 + m_t^2}$, where $\tilde{s}_\Delta^c$ is the lightest scalar. In the region $\Delta < \omega < \sqrt{ \Delta^2+ m_t^2}$, which is also shown in Fig.~\ref{fig:NeuBoundMax_Mix}, the lightest scalar is $\tilde{s}_\omega^c$. In this case we must exchange $m_{\tilde{s}_\Delta^c} \leftrightarrow m_{\tilde{s}_\omega^c}$, and make the replacements $\sin \phi_R \to \cos \phi_R$ and $\cos \phi_R \to -\sin \phi_R$.}
\begin{equation}
\Gamma(\chi_{\Delta \Delta} \to ij) = \frac{|\chi(0)|^2}{32\pi m_{\tilde{s}_{\Delta}^c}^2} \frac{\beta_{ij}}{1+\delta_{ij}} \sum|\mathcal{M}(ij)|^2\,,
\end{equation}
where $\sum|\mathcal{M}(ij)|^2$ and $\beta_{ij}$ are the spin-summed matrix element squared for $\tilde{s}_\Delta^{c\ast} \tilde{s}_\Delta^c \to ij$ and the final state velocity, respectively, both evaluated at threshold. The value of the wavefunction at the origin, $\chi(0)$, is unknown but does not affect the branching ratios. First of all, for the dominant $g_{B} g_{B}$ final state we have
\begin{equation}
\sum |\mathcal{M} ( g_{B} g_B)|^2 = 32 \pi^2 \frac{N_c^2 - 1}{N_c}  \alpha_d^2\,, 
\end{equation}
where $N_c = 3$. For $hh$ we find
\begin{equation}
 \sum |\mathcal{M} (hh)|^2 = N_c \left( \lambda_{hh\tilde{s}_\Delta^c \tilde{s}_\Delta^c} + \frac{3m_h^2 \kappa_{hhh} \lambda_{h\tilde{s}_\Delta^c \tilde{s}_\Delta^c} }{4m_{\tilde{s}_\Delta^c}^2-m_h^2} - 2\sum_{i\,=\,\Delta,\omega} \frac{\lambda_{h\tilde{s}_\Delta^c\tilde{s}_i^c}^2  v^2}{m_{\tilde{s}_\Delta^c}^2+m_{\tilde{s}_i^c}^2-m_h^2}\right)^2,
\end{equation}
where $\kappa_{hhh}=1$ and the couplings between the squirks and the Higgs read
\begin{equation}
\lambda_{hh\tilde{s}_\Delta^c \tilde{s}_\Delta^c}= y_t^2\cos^2\phi_R\,,\qquad \lambda_{h\tilde{s}_\Delta^c \tilde{s}_\Delta^c}= y_t^2  \cos^2\phi_R - \sqrt{2}\,  \frac{ y_t  \omega}{v} \sin\phi_R \cos\phi_R \,,
\end{equation}
\begin{equation}
\lambda_{h\tilde{s}_\Delta^c \tilde{s}_\omega^c}=  y_t^2   \sin\phi_R \cos\phi_R +  \frac{y_t \omega}{\sqrt{2}\,v}  (\cos^2\phi_R -  \sin^2\phi_R).
\end{equation}
For the $WW$ and $ZZ$ final states,
\begin{align}
\frac{4}{N_c g^4} \sum& |\mathcal{M} (W^+ W^-)|^2 = 2 (a_{VV}^T)^2 + \left[  4 m_{\tilde{s}_\Delta^c}^2 \Big(\tfrac{m_{\tilde{s}_\Delta^c}^2}{m_W^2} - 1\Big)   \,\tfrac{\kappa_{W\tilde{s}_\Delta^c \tilde{b}_B^{\prime c}}^2}{m_{\tilde{s}_\Delta^c}^2 + m_{\tilde{b}_B^{\prime c}}^2 -m_W^2} - \Big( \tfrac{2 m_{\tilde{s}_\Delta^c}^2}{m_W^2} - 1\Big) \,a_{VV}^T  \right]^2\,, \nonumber \\
\frac{4c_w^4}{N_c g^4}\sum& |\mathcal{M} (ZZ)|^2 = 2(a_{VV}^T)^2 + \bigg[  4 m_{\tilde{s}_\Delta^c}^2 \Big(\tfrac{m_{\tilde{s}_\Delta^c}^2}{m_Z^2} - 1\Big)\sum_{i\,=\,\Delta,\omega} \tfrac{\kappa_{Z\tilde{s}_\Delta^c \tilde{s}_i^c}^2}{m_{\tilde{s}_\Delta^c}^2 + m_{\tilde{s}_i^c}^2 -m_Z^2}  - \Big( \tfrac{2 m_{\tilde{s}_\Delta^c}^2}{m_Z^2} - 1\Big) a_{VV}^T \bigg]^2,
\end{align}
where we defined
\begin{equation}
a_{VV}^T =  \kappa_{VV \tilde{s}_\Delta^c \tilde{s}_\Delta^c} + \frac{\lambda_{h\tilde{s}_\Delta^c \tilde{s}_\Delta^c} v^2 \kappa_{hVV}}{4m_{\tilde{s}_\Delta^c}^2-m_h^2}\,,
\end{equation}
with $\kappa_{VV \tilde{s}_\Delta^c \tilde{s}_\Delta^c} = \kappa_{Z\tilde{s}_\Delta^c \tilde{s}_\Delta^c}=\sin^2 \phi_R$, $\kappa_{Z\tilde{s}_\Delta^c \tilde{s}_\omega^c}=\sin \phi_R \cos \phi_R$ and $\kappa_{hVV} = 1$. Finally for $t\bar{t}$ we find
\begin{equation}
\sum \left|\mathcal{M} (t\bar{t}\,)\right|^2 = 8 N_c^2 (m_{\tilde{s}^c_\Delta}^2 - m_t^2) \left(\frac{ m_t \kappa_{htt} \lambda_{h\tilde{s}^c_\Delta \tilde{s}^c_\Delta}}{4 m_{\tilde{s}^c_\Delta}^2 -m_h^2} \right)^2\,
\end{equation}
with $\kappa_{htt} = 1$.

\bibliography{fourDFSUSYbib}

\end{document}